\documentclass[11pt]{amsart}
\usepackage{epsfig}
\usepackage{verbatim}
\usepackage{graphicx,amssymb,amsmath,amsthm}
\usepackage{mathrsfs}
\usepackage{amssymb}
\usepackage{enumerate}
\newcommand{\lra}{{\longrightarrow}}
\newcommand{\eproof}{\hfill\rule{2.2mm}{3.0mm}}

\newcommand{\Proof}{\noindent {\bf Proof.~~}}

\renewcommand{\SS}{{\mathcal S}}

\newcommand{\R}{{\mathbb R}}

\newcommand{\C}{{\mathbb C}}

\newcommand{\N}{{\mathbb N}}
\newcommand{\PP}{{\mathbb P}}

\newcommand{\wmod}[1]{\mbox{~(mod~$#1$)}}
\renewcommand{\eqref}[1]{(\ref{#1})}
\newcommand{\inner}[1]{\langle #1 \rangle}
\newcommand{\abs}[1]{\lvert#1\rvert}
\newcommand{\shsp}{\hspace{1em}}
\newcommand{\mhsp}{\hspace{2em}}

\newcommand{\A}{{\mathcal A}}

\newcommand{\bI}{{\mathbf I}}
\newcommand{\m}{\mathfrak{m}}
\newcommand{\rank}{{\rm rank}}
\renewcommand{\span}{{\rm span}}
\newcommand{\diag}{{\rm diag}}
\newcommand{\tr}{{\rm Tr}}

\newcommand{\vx}{{\mathbf x}}
\newcommand{\vy}{{\mathbf y}}
\newcommand{\vz}{{\mathbf z}}
\newcommand{\vw}{{\mathbf w}}

\newcommand{\vv}{{\mathbf v}}
\newcommand{\vu}{{\mathbf u}}
\newcommand{\vf}{{\mathbf f}}

\renewcommand{\H}{{\mathbb F}}
\newcommand{\HH}{{\H}^d}
\newcommand{\qHH}{\underline{\HH}}
\newcommand{\ul}[1]{\underline{#1}}
\newcommand{\LL}{\mathbf L}
\newcommand{\MM}{\mathbf M}
\newcommand{\Herm}{{\mathbf H}_d}

\newcommand{\G}{{\mathcal G}}

\renewcommand{\top}{T}
\newtheorem{prop}{Proposition}[section]
\newtheorem{lem}[prop]{Lemma}
\newtheorem{defi}{Definition}[section]
\newtheorem{coro}[prop]{Corollary}
\newtheorem{theo}[prop]{Theorem}

\begin{document}
\baselineskip 18pt

%\title{Generalized Phase Retrieval and related problems}
\title[Generalized phase retrieval ]{Generalized phase retrieval : measurement number, matrix recovery and beyond }
\author{Yang Wang}
\thanks{Yang Wang was supported in part by the Hong Kong Research Grant Council grant 16306415.
% and the AFOSR grant FA9550-12-1-0455.
      Zhiqiang Xu was supported  by NSFC grant (11171336, 11422113, 11021101, 11331012) and by National Basic Research Program of China (973 Program 2015CB856000)}
\address{Department of Mathematics  \\ The Hong Kong University of Science and Technology\\
Clear Water Bay, Kowloon, Hong Kong}
\email{yangwang@ust.hk}
\author{Zhiqiang Xu}
\address{LSEC, Inst.~Comp.~Math., Academy of
Mathematics and System Science,  Chinese Academy of Sciences, Beijing, 100091, China}
\email{xuzq@lsec.cc.ac.cn}

\begin{abstract}
   In this paper, we develop  a  framework of generalized  phase retrieval in which one aims to reconstruct a vector $\vx$ in $\R^d$ or $\C^d$ through quadratic samples $\vx^*A_1\vx, \dots, \vx^*A_N\vx$. The generalized phase retrieval includes as special cases  the standard phase retrieval as well as the phase retrieval by orthogonal projections. We first explore the connections among generalized phase retrieval, low-rank matrix recovery and nonsingular bilinear form. Motivated by the connections,
   we present results on the minimal measurement number needed for recovering  a matrix
    that lies in a set $W\in \C^{d\times d}$. Applying the results to phase retrieval,
    we show that generic $d \times d$ matrices $A_1,\ldots, A_N$ have the phase retrieval
    property if $N\geq 2d-1$ in the real case and $N \geq 4d-4$ in the complex case for
     very general classes of $A_1,\ldots,A_N$, e.g. matrices with prescribed ranks or
      orthogonal projections.  Our method also leads to a novel  proof  for the
  classical Stiefel-Hopf condition on nonsingular bilinear form.  We also
give lower bounds on the minimal measurement number required for generalized
        phase retrieval. For several classes of dimensions $d$ we obtain the precise
        values of the minimal measurement number. Our work unifies and enhances results
        from the standard phase retrieval,  phase retrieval by projections and low-rank
         matrix   recovery.
%
%Our work unifies and enhances results from the standard phase retrieval,  phase retrieval by projections and low-rank matrix   recovery.    The proofs are often based on new ideas and techniques involving determinantal variety, topology and nonsingular bilinear form.
\end{abstract}
\maketitle

\section{Introduction}
\setcounter{equation}{0}

\subsection{Problem Setup}

The {\em phase retrieval } problem is to recover  signals from the magnitude of
the observations. It has important applications in imaging, optics, quantum tomography, communication,
audio signal processing and more, and it has grown into one of the major areas of research
in recent years (see e.g. \cite{BCE06,bodmann,CSV12,CEHV15,E15, FMW14, HMW13} and the references therein).
First we state the phase retrieval problem. In the finite dimensional Hilbert space $\H^d$, where $\H=\R$ or $\H=\C$,  a set of elements $\{\vf_1,\ldots,\vf_N\}$ in $\HH$ is called a {\em frame}
  if it spans $\HH$. Given this frame any vector $\vx\in {\HH}$ can be
   reconstructed from the inner products $\{\inner{\vx,\vf_1},\ldots,\inner{\vx,\vf_N}\}$. The {\em standard version} of the phase retrieval problem in $\HH$ is:
Let $\{\vf_1,\ldots,\vf_N\}$ be a subset in the finite
dimensional Hilbert space $\HH$. Is it possible to reconstruct a vector $\vx \in\HH$ from $\{\abs{\inner{\vx,\vf_1}},\ldots,\abs{\inner{\vx,\vf_N}}\}$, i.e. from
only the magnitude of the inner products? To do that, the set  $\{\vf_1,\ldots,\vf_N\}$ must be a frame because otherwise one can find a nonzero $\vx$ such that it is orthogonal to all $\vf_j, j=1,\ldots,N$. Furthermore if $\vx'= b\vx$ where $|b|=1$ then $|\inner{\vx,\vf_j}| = |\inner{\vx',\vf_j}|$ for all $j=1,\ldots,N$, and hence $\vx$ and $\vx'$ cannot be distinguished from the magnitude of the inner products. Thus all reconstructions from magnitudes, if it is possible, should only be up to a unimodular constant.

\subsubsection{Generalized Phase Retrieval}
There have been significant advances in the study of this standard version of the phase retrieval problem.
 On the one hand, many theoretical results are  presented. Particularly,   the problem of finding the minimal measurement number for phase retrieval has attracted a lot of attention  \cite{BCMN,BCE06,HMW13,CEHV15,V15, WaXu14}. On the other hand, efficient and numerically stable algorithms have been developed to solve for phase retrieval (see \cite{CSV12,CESV12}).

  In this paper, we focus on the more theoretical side of  a {\em generalized version} of the phase retrieval problem. The standard phase retrieval problem is to reconstruct a $\vx\in\H^d$ up to a unimodular constant from the measurements $\{\vx^*\vf_j\vf_j^*\vx=|\inner{\vx,\vf_j}|^2\}_{j=1}^N$. Set $A_j=\vf_j\vf_j^*$. Then the problem is to reconstruct $\vx$ from the measurements $\{\vx^*A_j\vx\}_{j=1}^N$, where $A_j$ are positive semidefinite and $\rank(A_j)=1$. In the generalized phase retrieval problem, the restrictions on $A_j$ are relaxed and replaced, and one aims to reconstruct $\vx$ up to a unimodular constant from more general {\em quadratic measurements}  $\{\vx^*A_j\vx\}_{j=1}^N$.

  Let $\Herm(\H)$ denote the set of $d\times d$ Hermitian matrices over $\H$ (if $\H=\R$ then Hermitian matrices are symmetric matrices). As with the standard phase retrieval problem we consider the equivalence relation $\sim$ on $\HH$: $\vx_1 \sim \vx_2$ if there is a constant $b\in \H$ with $|b|=1$ such that $\vx_1=b\vx_2$. Let $\qHH :=\HH/\sim$. We shall use $\ul\vx$ to denote the equivalent class containing $\vx$. For any given $\A=(A_j)_{j=1}^N \subset \Herm(\H)$ define the map $\MM_\A: \qHH \lra \R^N$ by
\begin{equation}  \label{1.1}
      \MM_\A(\ul\vx) = (\vx^*A_1\vx, \dots, \vx^*A_N\vx).%^\top.
\end{equation}
Thus the {\em generalized phase retrieval problem} asks whether we can reconstruct $\ul\vx\in \qHH$ from $\MM_\A(\ul\vx)$. We should observe that $\MM_\A$ can also be viewed as a map from $\HH$ to $\R^N$, and we shall often do this when there is no confusion.

\begin{defi}  \label{defi-1.1}{\rm
    Let $\A=(A_j)_{j=1}^N \subset \Herm^N(\H)$. We say $\A$ has the {\em phase retrieval property} or is {\em phase retrievable (PR)} if $\MM_\A$ is injective on $\qHH$.
}
\end{defi}

Note that the generalized phase retrieval problem includes the standard phase retrieval problem as a special case, with the additional restrictions $A_j \succeq 0$ and $\rank(A_j)=1$. It also includes the so-called {\em fusion frame (or projection) phase retrieval} as a special case where each $A_j$ is an orthogonal projection matrix, namely $A_j^2=A_j$ \cite{E15,phaseproj1,phaseproj2}. Moreover, it is very closely related to and a generalization of the problem of information completeness of positive operator valued measures (POVMs) with respect to pure states in quantum tomography \cite{HMW13}, where the norm of the vector we try to recover $\vx\in\C^d$ is assumed to be 1. So in essence  information completeness of POVMs with respect to pure states is a special case of generalized phase retrieval in $\C^d$ in which one of the measurement matrix $A_j$ is the identity matrix $I_d$.
The generalized phase retrieval problem, just like the standard phase retrieval problem, has in fact several flavors involving different subtleties, some of which will be discussed later in the paper. One of the most fascinating aspect of  generalized phase retrieval is its close connections to other areas in mathematics,  which  include matrix recovery, nonsingular bilinear form,  composition of quadratic forms and the embedding problem in topology.

This paper attempts to lay down a foundation for generalized phase retrieval by establishing several fundamental properties. Of particular interest is the various minimality problems for generalized phase retrieval, and its connections to matrix  recovery and nonsingular bilinear form. We list some of them below:

\vspace{2mm}
\noindent
{\bf Minimality Questions for Generalized Phase Retrieval:}~~{\em Let $\A=(A_j)_{j=1}^N \subset \Herm^N(\H)$. What is the smallest $N$ so that a generic $\A=(A_j)_{j=1}^N$ has the phase retrieval property in $\H^d$?}

\vspace{2mm}

There can also be numerous variants of those aforementioned questions. For example, what if we require that all $A_j \succeq 0$? What if we prescribe the ranks for all $A_j$? We can obviously impose various special restrictions on $A_j$, and any such restrictions may alter the answer to each of the above questions.

\subsubsection{Generalized Matrix Recovery}
Note that $\vx^*A_j\vx={\rm Tr}(A_j\vx\vx^*)$. The generalized phase retrieval problem is  equivalent to the recovery of  the rank one Hermitian matrix  $\vx\vx^*$ from $({\rm Tr}(A_1\vx\vx^*),\ldots,{\rm Tr}(A_N\vx\vx^*))$, which establishes a natural connection between generalized phase retrieval and low-rank matrix recovery. The connection is observed in \cite{CSV12} and  Cand\`{e}s, Strohmer and Voroninski use it to study the standard phase retrieval. This method is called {\em PhaseLift}.

  The {\em low-rank matrix recovery }  problem is an active research area in recent years and has arisen in many important applications such as  image processing,  recommender systems and Euclidean embedding and more.
The goal of low-rank matrix recovery is to recover $Q\in \C^{d\times  d}$ with ${\rm rank}(Q)\leq r$ from linear observation $({\rm Tr}(A_1Q),\ldots,{\rm Tr}(A_NQ))\in \H^N$ for some given $A_1, \dots, A_N$. Depending on the problem and application, one  imposes various special restrictions on $A_j$ and $Q$, e.g.  all matrices  $A_1,\ldots, A_N$ have rank one \cite{ROP, ROP1}, and/or some of entries of $Q$ are $0$ etc. The generalized phase retrieval leads us naturally to the following generalized  matrix recovery problem:

\vspace{3mm}
\noindent
{\bf Generalized Matrix Recovery Problem:}~~{\em
Let $L:\H^{d\times d}\times \H^{d\times d}\rightarrow \H$ be a bilinear function. Let  $W\subset \H^{d\times d}$ and $V_j\subset \H^{d\times d}$ for $ j=1,\ldots,N$. Assume that $\A=(A_j)_{j=1}^N$ with $A_j\in V_j$.
Can we reconstruct any $Q\in W$   from $ \MM_\A(Q):=(L(A_1,Q),\ldots,L(A_N,Q))\in \H^N$?}

\vspace{3mm}

In this paper,  the sets $V_j$ and $W$ above will be taken to be algebraic varieties in $\H^{d\times d}$. We  also require that $W-W\subset \H^{d\times d}$ is  an algebraic variety, where
\[
W-W\,\,:=\,\, \{ \vx-\vy : \text{ for all } \vx, \vy \in W\}.
\]
Low-rank matrix recovery under different  conditions  usually becomes a  special cases of  the generalized matrix recovery problem in this setting. We list some examples here:

\begin{itemize}
 \item Let
$$
  {\mathcal M}_{d,r}(\H):=\Bigl\{Q\in \H^{d\times d}: {\rm rank}(Q)\leq r\Bigr\}, \mhsp\H=\C \mbox{~or~}\R.
$$
Note that ${\rm rank}(Q)\leq r$ is equivalent to the vanishing  of all $(r+1)\times (r+1)$ minors of $Q$ and that these $(r+1)\times (r+1)$ minors are  homogeneous polynomials in the entries of $Q$.
Hence, ${\mathcal M}_{d,r}(\H)$ is an algebraic variety in $\H^{d\times d}$.
If we take $W={\mathcal M}_{d,r}(\H)$, then the generalized matrix recovery problem is the rank $r$ matrix recovery problem.

\item If $V_j$ is the algebraic variety  containing matrices of rank $\leq 1$ then matrix recovery problem becomes the problem of  matrix recovery by rank one projections \cite{ROP}.

\item An interesting and important problem is the recovery of low-rank sparse matrices. Set
$$
  {\Sigma}_{d,k}(\H):=\Bigl\{Q\in \H^{d\times d} : \|Q\|_0\leq k \Bigr\}, \mhsp\H=\C \mbox{~or~}\R,
$$
where $\|Q\|_0$ denotes the nonzero entries of $Q$. Then $Q\in \Sigma_{d,k}$ if and only if the product of any $k+1$ entries in $Q$ vanishes which implies $\Sigma_{d,k}$ is an algebraic variety.    Thus the recovery of sparse matrices is a special case of generalized matrix recovery by taking $W=\Sigma_{d,k}(\H)$ or $W=\Sigma_{d,k}(\H)\cap   {\mathcal M}_{d,r}(\H)$.

\item We often meet the case where the measurement matrix  is a  Hermite matrix. The Hermite matrix set $\Herm(\C)$ is {\em not} an algebraic variety but we can transform it to the setting with  $V_j=\R^{d\times d}$ by choosing
    an appropriate bilinear function $L$. Define a linear map $\tau: \C^{d\times d} \lra \C^{d\times d}$ by
$$
    \tau(A) = \frac{1}{2} (A+A^\top) +\frac{i}{2} (A-A^\top).
$$
It is easy to see that $\tau$ restricted on $\R^{d\times d}$ is an isomorphism from $\R^{d\times d}$ to $\Herm(\C)$.
Set $L(A_j,Q):=\tr(\tau(A_j)Q)$. Then we can take $V_j=\R^{d\times d}$ which is a real algebraic variety.
\end{itemize}

\vspace{3mm}
%For generalized matrices recovery, we focus on the following question in this paper:

%\vspace{1mm}
\noindent
{\bf Minimality Question for Generalized Matrix Recovery:}~~{\em Let $L:\H^{d\times d}\times \H^{d\times d}\rightarrow \H$ be a bilinear form. Let $V_j\subset \H^{d\times d}$ for $ j=1,\ldots,N$  and $W\subset \H^{d\times d}$ be algebraic varieties. Assume that $\A=(A_j)_{j=1}^N$ with $A_j\in V_j$. Under what conditions can we reconstruct any $Q\in W$   from $ \MM_\A(Q):=(L(A_1,Q),\ldots,L(A_N,Q))\in \H^N$? In particular, what is the smallest $N$ so that $\MM_\A$ is injective on $W$ for  a generic $\A=(A_j)_{j=1}^N\in V_1\times\cdots \times V_N$?}
\vspace{3mm}

Note that $\MM_\A$ is injective on $W$ if and only if, for  $Q\in W-W$, $\MM_\A(Q)=0$ implies that $Q=0$.  Throughout the rest of this paper, to state conveniently,  we abuse the notations and still use $W$ to denote $W-W$.  We  will employ algebraic method to  investigate the smallest $N$ so that $\{Q\in W: \MM_\A(Q)=0\}$ only contains the zero point which implies the answer for the question above. The results  will play an important role in generalized phase retrieval.

\subsection{Related Results }

\subsubsection{ Phase Retrieval  and Matrix Recovery}

For the standard phase retrieval with $\H=\R$ the minimality question is relatively straightforward. Let $\A=(A_j)_{j=1}^N \subset \Herm(\R)$ such that $A_j = \vf_j \vf_j^*$ for some $\vf_j\in \R^d$. Then it is easy to prove that the smallest $N$ for which $\A$ can have the phase retrieval property  is $N=2d-1$, which is also the smallest number that a generic such $\A$ with $N$ elements has the phase retrieval property \cite{BCE06}. However, once we remove the $\rank(A_j) = 1$ condition the answers are already different. For example for fusion frame phase retrieval in $\R^d$, it is known that a generic choice of $N=2d-1$ orthogonal projections $\A=(P_j)_{j=1}^N$ with $0<\rank(P_j)<d$ has the phase retrieval property \cite{phaseproj1,E15}, but the smallest such $N$ remains unknown in general. For $d=4$, it is known that there exists a fusion frame $\A=(P_j)_{j=1}^N$ with $N=6 = 2d-2$ \cite{Xu15}  having the phase retrieval property. In this paper, we shall show the number $N=6$ is tight for $d=4$.

In the complex case $\H=\C$, the same question remains open for the standard phase retrieval. It is known that in the standard phase retrieval setting, $N\geq 4d-4$ generic matrices $\A=(A_j)_{j=1}^N \subset \Herm(\C)$ where $A_j=\vf_j\vf_j^*$ have the phase retrieval property \cite{BCMN, CEHV15}. Moreover, the $N=4d-4$ is also minimal if $d=2^k+1$ where $k \geq 1$ \cite{CEHV15}.  Vinzant in \cite{V15} has constructed an example in $d=4$ with $N=11 =4d-5<4d-4$ matrices $A_j=\vf_j\vf_j^*$ such that $\A=(A_j)_{j=1}^{11}$ is phase retrievable in $\C^4$. The construction is done through the use of computational algebra tools and packages. This result implies that $N=4d-4$ is not minimal for some $d$ for the standard phase retrieval. So far, the smallest $N$  is not known even for $d=4$. In the other direction, a lower bound  $N\geq 4d-3-2\alpha$ for the minimal $N$ is given in \cite{HMW13}, where $\alpha$ denotes the number of $1$'s in the binary expansion of $d-1$. This was the best known lower bound for standard phase retrieval.

Recall that we use ${\mathcal M}_{d,r}(\H)$ to denote the set of $d\times d$ matrices in $\H^{d\times d}$ with rank $\leq r$.  For low-rank matrix recovery, any $Q\in {\mathcal M}_{d,r}(\H)$ can be recovered from $(\tr(A_jQ))_{j=1}^N$ with probability 1 if $N \geq 4dr-4r^2$, where the matrices $A_1,\ldots, A_N$ are i.i.d. Gaussian random matrices, provided  $r \leq d/2$.  It was also conjectured in \cite{uniq} that $N=4dr-4r^2$ is the minimal $N$ for which there exists $\A=(A_j)_{j=1}^N$ so that $\MM_\A$ is injective on  ${\mathcal M}_{d,r}(\H)$. In \cite{Xu15}, the author proved the conjecture for $\H=\C$ and disproved it for $\H=\R$, showing the existence of $\A=(A_j)_{j=1}^{11}$ for which $\MM_\A$ is injective on ${\mathcal M}_{4,1}(\R)$.

\subsubsection{Nonsingular Bilinear Form}
As we will show in Theorem \ref{prop-2.1}, $\A=(A_j)_{j=1}^N\subset \Herm^N(\R)$ having the phase retrieval property is equivalent to the corresponding bilinear form  $(\vx^\top A_j\vy)_{j=1}^N$ being nonsingular. This connection has led us to also study nonsingular bilinear form, an area with deep historical roots.  Consider the bilinear form $\LL:\R^p\times \R^q\rightarrow \R^N$  given by $\LL(\vx,\vy)=(\vx^\top B_1\vy,\ldots,\vx^\top B_N\vy)\in \R^N$ where $\vx\in \R^p, \vy \in \R^q$ and $B_j\in \R^{p\times q}$. We shall call $(p,q,N)$ the {\em size} of  $\LL$. The bilinear form is {\em nonsingular} if $\LL(\vx,\vy)=0$ implies $\vx=0$ or $\vy=0$; it is {\em normed} if $\abs{\LL(\vx,\vy)}=\abs{\vx} \cdot \abs{\vy}$. A simple observation is that if $\LL$ is normed then it is nonsingular. We use $p\# q$ to denote the minimal $N$ for which there exist $B_1,\ldots,B_N$ such that the corresponding bilinear form is nonsingular. The function $p\# q$ appears in the study of the composition of quadratic forms and the immersion problem \cite{surv, Shapirobook}.  It is well-known that $2\#2=2$. In 1748, Euler found a normed bilinear form with size $(4,4,4)$ in his attempt to prove Fermat's Last Theorem \cite{Shapirobook}, which implies $4\#4=4$.  Degen  proved $8\# 8=8$ in 1818.   The exact values of $p\# q$ for some small $p,q\leq 32$ are known and can be found in \cite{Shapirobook}. However, finding the exact value for  $p\# q$ in general is a very hard problem.  A well-known necessary condition for the existence of a nonsingular bilinear form of size $(p,q,N)$ is the Stiefel-Hopf condition, proved by Hopf and Stiefel independently in 1941 (see also \cite{hopf,Lam}).

\begin{theo}(Stiefel-Hopf)\label{th:hopf}
 If there exists a nonsingular bilinear form of size $(p,q,N)$ then the binomial coefficient $\binom{N}{k}$ is even whenever $N-q+1\leq k\leq p-1$.
\end{theo}

In the generalized phase retrieval setting it always requires $p=q=d$ together with the additional requirement that matrices $B_j, j=1,\ldots,N,$ are symmetric.  Thus for our study we are interested in the minimal $N$ for which there exists a nonsingular symmetric bilinear form of size $(d,d,N)$. This is a stronger requirement so $N\geq d\# d$ and $(d,d,N)$ should satisfy the Stiefel-Hopf condition.

\subsection{Our Contribution}
Our study focuses on the number of measurements needed to achieve generalized phase retrieval
 and other related questions. For these purposes we use the notation $\m_\H(d)$ to denote
  minimal $N$ for which phase retrieval property is possible:
$$
   \m_\H(d)\,:=\,\min\Bigl\{N: ~\text{ there exists  a phase retrievable $\A=(A_j)_{j=1}^N\subset \Herm^N(\H)$ in~}\H^d\Bigr\}.
$$
We use algebraic methods to study the measurement number $N$ for which a generic $\A=(A_j)_{j=1}^N$ has the phase retrieval property. We also present an upper bound for $\m_\H(d)$. Meanwhile a lower bound for $\m_\H(d)$ is obtained using results on the embedding of projective spaces into real spaces.  These results also show a direct link among phase retrieval, matrix recovery and  nonsingular bilinear form. In Section 2, we give several equivalent formulations for generalized phase retrieval, where we establish its close connection to nonsingular bilinear form and matrix recovery.
In Section 3, we investigate the number of  measurements needed for generalized matrix recovery, by showing that $N=\dim (W)$ measurements are necessary, and moreover sufficient for generic measurements in the case $\H=\C$ provided the algebraic varieties $V_j, j=1,\ldots,N$ and $W$ satisfy some mild conditions.   The tools from algebraic geometry play an important role in our investigation. Using these tools we also provide an alternative proof for the Stiefel-Hopf condition (Theorem \ref{th:hopf}), which may be independently interesting in itself.  In Section 4 we show  that $N=2d-1$ (resp. $N=4d-4$) generic matrices with prescribed ranks have the phase retrieval property in $\R^d$ (resp. $\C^d$). Similar technique also allows us to establish the $N=4d-4$ result for generic fusion frames, namely $N=4d-4$ generic orthogonal projections have the phase retrieval property in $\C^d$. Finally, in Section 5, we study the minimal measurement number $\m_\H(d)$ by employing the results on the embedding of projective spaces in Euclidean spaces. In the real case $\H=\R$, we prove that $2d-O(\log_2d) \leq \m_\R(d)\leq 2d-1$. When $d$ is of the form $d=2^k+\delta$ where $\delta=1$ or $2$, we obtain the exact value $\m_\R(d)=2d-\delta$.
In the complex case $\H=\C$, let $\alpha$ denotes the number of 1's in the binary expansion of $d-1$. Then the lower bound $4d-2-2\alpha $ was obtained for information completeness of POVMs with respect to pure states \cite{HMW13}, which leads to the lower bound $ 4d-3-2\alpha $ for the  phase retrieval. In this paper we improves the results to  $\m_\C(d)\geq 4d-2-2\alpha $. As a result, combining with known upper bounds we are able to obtain the exact value of $\m_\C(d)$ for several classes of dimensions $d$, including particularly the special case  $d=2^k+1>4$, for which $\m_\C(d)=4d-4$. This sharp lower bound in the standard phase retrieval setting was first shown in  \cite{CEHV15}.

\section{Equivalent Formulations for Generalized Phase Retrieval}
\setcounter{equation}{0}

We state an equivalent formulation for the generalized phase retrieval problem here, which allows us to prove some basic but important properties for generalized phase retrieval.

For any $c\in\C$ let $\Re(c)$ and $\Im(c)$ denote the real and imaginary part of $c$, respectively. A useful formula is that for a Hermitian $A\in \Herm(\H)$ and any $\vx,\vy\in\H^d$ we must have
\begin{equation} \label{2.1}
       \vx^* A\vx - \vy^*A\vy = 2\Re (\vv^*A\vu)
\end{equation}
where $\vv=\frac{1}{2}(\vx+\vy)$  and $\vu=\frac{1}{2}(\vx-\vy)$. This is straightforward to check. In the real case $\H=\R$ it means that $\vx^* A\vx - \vy^*A\vy = \vv^*A\vu =\vv^\top A\vu$.

\begin{theo}  \label{prop-2.1}
   Let $\A=(A_j)_{j=1}^N \subset \Herm(\R)$. The following are equivalent:
   \begin{itemize}
   \item[\rm (1)] $\A$ has the phase retrieval property.
   \item[\rm (2)] There exist no nonzero $\vv,\vu\in\R^d$ such that $\vv^\top A_j \vu=0$ for all $1\leq j\leq N$.
   \item[\rm (3)] $\span\{A_j \vu\}_{j=1}^N=\R^d$ for any nonzero $\vu\in\R^d$.
   \item[\rm (4)] If $Q\in {\mathcal M}_{d,1}(\R)$ and $\tr(A_jQ)=0$ for all $1 \leq j \leq N$, then $Q=0$.
   \item[\rm (5)] For any nonzero $Q\in {\mathcal M}_{d,2}(\R)\cap \Herm(\R)$ such that $\tr(A_jQ)=0$ for all $1 \leq j \leq N$, $Q$ has two nonzero eigenvalues having the same sign.
   \item[\rm (6)] The bilinear form $\LL: \R^d \times \R^d \lra \R^N$ given by $\LL(\vx,\vy):=(\vx^\top A_j\vy)_{j=1}^N$ is nonsingular.
   \item[\rm (7)] The Jacobian of $\MM_\A$ has rank $d$ everywhere on $\R^d\setminus\{0\}$.
\end{itemize}
\end{theo}
\Proof  (1) $\Leftrightarrow$ (2). ~This is rather clear. If there exist $\vx \neq \pm \vy$ in $\R^d$ such that $\MM_\A(\vx)=\MM_\A(\vy)=0$ then $\vx^\top A_j\vx -\vy^\top A_j\vy=(\vv+\vu)^\top A_j(\vv+\vu)-(\vv-\vu)^\top A_j(\vv-\vu)=0$ which implies $\vv^\top A_j\vu=0$ for all $j$, where $\vv=\frac{1}{2}(\vx+\vy)$  and $\vu=\frac{1}{2}(\vx-\vy)$. Clearly, both $\vu, \vv$ are nonzero. This is a contradiction. The converse also follows from the same argument.

(1) $\Leftrightarrow$ (5). ~We first show (1) $\Rightarrow$ (5) by contradiction. Assume there is a $Q\in {\mathcal M}_{d,2}(\R)\cap \Herm(\R)$ such that $\tr(A_jQ)=0$ for all $j$ and $Q$ has two nonzero eigenvalues $\lambda_1>0$ and $\lambda_2<0$. By spectral decomposition we can write $Q$ as
$$
 Q =\lambda_1 \vu \vu^\top-\abs{\lambda_2} \vv \vv^\top
$$
where $\inner{\vu,\vv}=0$. Thus
$$
    \tr(A_j(\lambda_1 \vu \vu^\top-\abs{\lambda_2} \vv \vv^\top))
        =\tr(A_j \vx \vx^\top)-\tr(A_j \vy \vy^\top)=0
$$
where $\vx=\sqrt{\lambda_1}\vu, \vy=\sqrt{\abs{\lambda_2}}\vv$.
Since $\vx^\top A_j\vx={\rm Tr}(A_j\vx\vx^\top )$ and $\vy^\top A_j\vy={\rm Tr}(A_j\vy\vy^\top )$, it follows that $\MM_\A(\vx)=\MM_\A(\vy)$. But $\vx\neq \pm \vy$, this contradicts with (1).

We next show (5) $\Rightarrow$ (1). Assume there exist $\vx, \vy \in \R^d$ so that $\vx^\top A_j\vx=\vy^\top A_j\vy$ for all $j$ and $\vx\neq \pm \vy$. Then
$$
{\rm Tr}(A_j (\vx \vx^\top-\vy \vy^\top))=\vx^\top A_j\vx-\vy^\top A_j\vy=0.
$$
Set $Q:=\vx\vx^\top -\vy\vy^\top \neq 0$. Then $Q\in {\mathcal M}_{d,2}(\R)\cap \Herm(\R)$ such that $\tr(A_jQ)=0$ for all $j$. Hence $Q$ has two nonzero eigenvalues of the same sign. This implies that $\vx$ and $\vy$ are linearly independent, and therefore $Q$ has two nonzero eigenvalues with opposite signs,  contradicting (5).

(2) $\Leftrightarrow$ (3). ~If for some nonzero $\vu_0\in \R^d$ so that ${\rm span}\{A_j\vu_0\}_{j=1}^N\neq \R^d$. Then we can find $\vv_0\neq 0$ so that $\vv_0\bot {\rm span}\{A_j\vu_0\}_{j=1}^N$. This implies $\vv_0^\top A_j\vu_0=0$ for all $j$. The converse is clearly also true from the same argument.

(2) $\Leftrightarrow$ (6). ~The bilinear form $\LL$ is nonsingular if and only if $\LL(\vx,\vy) \neq 0$ for all nonzero $\vx,\vy$. This is precisely the condition in (2).

(4) $\Leftrightarrow$ (6). ~First we observe that $Q \in  {\mathcal M}_{d,1}(\R)$ if and only if $Q=\vx \vy^\top$, and $Q\neq 0$ if and only if both $\vx,\vy \neq 0$. The equivalence follows immediately from the fact $\LL(\vx,\vy)= (\tr(A_jQ)_{j=1}^N$ where $Q=\vx \vy^\top$.

(3) $\Leftrightarrow$ (7). ~The Jacobian of $\MM_\A$ at $\vx$ is exactly $J_\A(\vx)=2[A_1\vx, A_2\vx, \dots, A_N\vx]$, i.e. the columns of $J_\A(\vx)$ are precisely $\{A_j\vx\}_{j=1}^N$. Thus (3) is equivalent to for any $\vx\neq 0$ the rank of $J(\vx)$ is $d$.
\eproof

\medskip

We remark that the equivalence of some of these conditions are known for the standard phase retrieval. The equivalence of (3) and (1) was also established for real orthogonal projections matrices in \cite{E15}.

\begin{theo}  \label{prop-2.2}
   Let $\A=(A_j)_{j=1}^N \subset \Herm(\C)$. The following are equivalent:
   \begin{itemize}
   \item[\rm (1)] $\A$ has the phase retrieval property.
   \item[\rm (2)] There exist no $\vv,\vu\neq 0$ in $\C^d$ with $\vu \neq ic\vv$ for any $c\in\R$ such that $\Re(\vv^* A_j \vu)=0$ for all $1\leq j\leq N$.
   \item[\rm (3)] The (real) Jacobian of $\MM_\A$ has (real) rank $2d-1$ everywhere on $\C^d \setminus\{0\}$.
   \item[\rm (4)] For any nonzero $Q\in {\mathcal M}_{d,2}(\C)\cap \Herm(\C)$ such that $\tr(A_jQ)=0$ for all $1 \leq j \leq N$, $Q$ has two nonzero eigenvalues having the same sign.
\end{itemize}
\end{theo}
\Proof  (1) $\Leftrightarrow$ (2). ~Assume that there exist $\vv,\vu\neq 0$ in $\C^d$, $\vu \neq ic\vv$ for some $c\in\R$ such that $\Re(\vv^* A_j \vu)=0$ for all $1\leq j\leq N$. Set $\vx=\vu+\vv$ and $\vy = \vu-\vv$. We have $\MM_\A(\vx)=\MM_\A(\vy)$ by (\ref{2.1}). We show $\vx \neq a\vy$ whenever $|a|=1$. If otherwise, note that $a \neq \pm 1$ because $\vu, \vv \neq 0$. Hence we must have $\vu = \frac{a+1}{a-1}\vv$. But $\frac{a+1}{a-1}$ is pure imaginary, which is a contradiction. Thus $\MM_\A$ is not injective on $\qHH=\C^d/\sim$ and $\A$ is not phase retrievable.

Conversely assume that $\MM_\A$ is not injective and $\MM_\A(\vx)=\MM_\A(\vy)$ where $\vx \neq a\vy$ for $|a|=1$. Set $\vu = \vx+\vy$ and $\vv=\vx-\vy$. Then $\vu \neq ic \vv$ for any $c\in\R$. Furthermore, $\Re(\vv^* A_j \vu)=0$ for all $1\leq j\leq N$.

%~ We first show that (1) $\Longrightarrow$ (2). For the aim of contradiction, assume that there exist $\vv,\vu\neq 0$ in $\C^d$, $\vu \neq ic\vv$ for any $c\in\R$ such that $\Re(\vv^* A_j \vu)=0$ for all $1\leq j\leq N$. Set $\vx=\vu+\vv$ and $\vy = \vu-\vv$. We have $\MM_\A(\vx)=\MM_\A(\vy)$ by (\ref{2.1}). We claim $\vx \neq a\vy$ whenever $|a|=1$ which contradicts with (1). Hence, (2) holds. We remain to show that $\vx \neq a\vy$ whenever $|a|=1$. If otherwise, note that $a \neq \pm 1$ because $\vu, \vv \neq 0$. Hence we must have $\vu = \frac{a+1}{a-1}\vv$ but $\frac{a+1}{a-1}$ is pure imaginary, which contradicts with the assumption  of
% $\vu \neq ic\vv$ where $c\in\R$.
%We next show that (2) $\Longrightarrow$ (1).
%Assume that $\MM_\A$ is not injective, i.e., there exist $\vx,\vy\in \C^d$ with  $\vx \neq a\vy$ for $|a|=1$  so that $\MM_\A(\vx)=\MM_\A(\vy)$. Set $\vu = \vx+\vy$ and $\vv=\vx-\vy$. Then $\vu \neq ic \vv$ for any $c\in\R$ while  $\Re(\vv^* A_j \vu)=0$ for all $1\leq j\leq N$ which contradicts with (2).

(1) $\Leftrightarrow$ (4). ~The proof is almost identical to the proof of the equivalence of (1) and (5) in Theorem \ref{prop-2.1}. We omit the detail here.

(2) $\Leftrightarrow$ (3). ~Write $A_j = B_j+iC_j$ where $B_j, C_j$ are real. Then $B_j^\top =B_j$ and $C_j^\top =-C_j$. Let
\begin{equation}  \label{2.2}
     F_j = \begin{bmatrix}
         B_j & -C_j\\C_j & B_j
     \end{bmatrix}.
\end{equation}
Then for any $\vu = \vu_R + i\vu_I\in\C^d$ we have $\vu^*A_j\vu =\vx^\top  F_j \vx$, where $\vx^\top  = [\vu_R^\top , \vu_I^\top ]$. Thus the real Jacobian of $\MM_\A(\vu)$ is precisely
$$
     J_\A(\vu)= 2[F_1\vx, F_2\vx, \dots, F_N\vx].
$$
Note that
$$
   [-\vu_I^\top , \vu_R^\top ]F_j\vu = -\vu_I^\top B_j\vu_R+\vu_R^\top C_j\vu_R +\vu_I C_j \vu_I +\vu_R^\top B_j\vu_I = 0.
$$
Thus the rank of $J_\A(\vu)$ is at most $2d-1$. Moreover, for any $\vv = \vv_R+i \vv_I\in\C^d$ we have
$$
     2[\Re(\vv^*A_j\vu)] = [\vv_R^\top , \vv_I^\top ]J_\A(\vu).
$$

To prove (2) implies (3), assume there exist nonzero $\vu,\vv\in\C^d$ with $\vu \neq ic\vv$ for any $c\in\R$ such that $\Re(\vv^* A_j \vu)=0$ for all $1\leq j\leq N$. Denote $\vx^\top  = [\vu_R^\top , \vu_I^\top ]$ and $\vy^\top  = [\vv_R^\top , \vv_I^\top ]$. Then $\vy^\top  F_j \vx=0$ for all $j$. But $\vu \neq ic\vv$ implies $\vy^\top  \neq c [-\vu_I^\top , \vu_R^\top ]$ for any real $c$. Hence the rank of $J_\A(\vu)$ is at most $2d-2$.

Conversely, to prove (3) implies (2), assume there exists a nonzero $\vu\in\C^d$ such that the rank of $J_\A(\vu)$ is at most $2d-2$ then we can find a $\vy\in\R^{2d}$ such that $\vy^\top  J_\A(\vu)=0$ and $\vy^\top $ is not co-linear with $ [-\vu_I^\top , \vu_R^\top ]$. Write $\vy^\top  = [\vv_R^\top , \vv_I^\top ]$ and $\vv = \vv_R + i\vv_I$. Then $\vv \neq ic\vu$, and moreover $\Re(\vv^* A_j \vu)=0$ for all $j$.
\eproof

\medskip

In the standard phase retrieval, the set of the frames $(\vf_1,\ldots,\vf_N)\in \C^{d\times N} $ having the phase retrieval property in $\C^d$ is an open set \cite{Ra13, CEHV15}.
%Viewing $\A:=[A_1,\ldots,A_N]$ that is phase retrieval on $\H^d$ as a subset of $\Herm^N(\H):=\underbrace{\Herm(\H)\times \cdots \Herm(\H)}_N$.
The conclusion also holds for generalized  phase retrieval.

\begin{theo} \label{theo-2.3}
Let $\H=\R$ or $\C$. For any given $N$, the set of  $\A:=(A_j)_{j=1}^N\subset  \Herm^N(\H)$ having the phase retrieval property is an open set in $\Herm^N(\H)$.
\end{theo}
\Proof  We only need to prove that the set of $\A$'s not having the phase retrieval property is closed. First we consider the real case $\H=\R$.  Let $\{\A_n\}\subset \Herm^N(\H)$ be a sequence of $N$-tuples of real symmetric matrices that do not have the phase retrieval property and $\lim_n \A_n =\A$. By Theorem \ref{prop-2.1} there exists a $\vx_n\in \R^d\setminus\{0\}$ such that the Jacobian has $\rank J_{\A_n}(\vx_n) <d$ for any $n$. Without loss of generality we may assume $\|\vx_n\|=1$. Thus there is a subsequence $\vx_{n_k} $ with $\lim_k \vx_{n_k} = \vx$. Clearly $\|\vx\|=1$. Furthermore, $J_{\A_{n_k}}(\vx_{n_k}) \lra J_\A(\vx)$ and therefore $\rank J_{\A}(\vx) <d$. Thus $\A$ does not have the phase retrieval property, which proves that the set of all non-phase retrieval $\A$'s is closed. This yields the theorem for $\H=\R$.

For the complex case $\H=\C$ the proof is essentially identical. Let $\A_n$ be a sequence of $N$-tuples of Hermitian matrices that do not have the phase retrieval property and $\lim_n \A_n =\A$. By Theorem \ref{prop-2.2} there exists a nonzero $\vu_n \in\C^d$ such that the real Jacobian $J_{\A_n}(\vu_n)$ has rank at most $2d-2$. Without loss of generality we may assume $\|\vu_n\|=1$. Thus there is a subsequence $\vu_{n_k} $ with $\lim_k \vu_{n_k} = \vu$. Clearly $\|\vu\|=1$. Furthermore, $J_{\A_{n_k}}(\vu_{n_k}) \lra J_\A(\vu)$ and hence $\rank J_{\A}(\vu) \leq 2d-2$. Thus $\A$ does not have the phase retrieval property, which proves that the set of all non-PR $\A$'s is closed. This yields the theorem for $\H=\C$.
\eproof

Theorem \ref{theo-2.3} implies the following Corollary:

\begin{coro}   \label{coro-2.4}
   The phase retrieval property over $\H$ for $\A \in \Herm^N(\H)$ is preserved under small perturbation.
\end{coro}

\section{ The Generalized Matrix Recovery and Nonsingular  Bilinear Form}
\setcounter{equation}{0}

In this section we present results on the recovery of  matrices. We establish its connection to phase retrieval, and use it to investigate nonsingular bilinear form. The main result here serves as the foundation of our results on generalized phase retrieval.

\subsection{Terminology From Algebraic Geometry}
 We first introduce some basic notations and results from algebraic geometry that are useful for this paper. Let $V \subseteq\C^d$ be an algebraic variety, i.e. $V$ is the locus of a collection of polynomials in $\C[\vx]$. We shall use $\bI(V)$ to denote the ideal of $V$, i.e.,
$$
      \bI(V) :=\Bigl\{f\in\C[\vx]:~ f\equiv 0 ~\mbox{on $V$}\Bigr\}.
$$
The ideal $\bI(V)$ is always a finitely generate radical ideal. We write $\bI(V)=\left< g_1, \dots, g_m\right>$ to denote that $\bI(V)$ is generated by the polynomials $g_1, \dots, g_m\in \C[\vx]$. It is well known that there is a one-to-one correspondence between radical ideals of $\C[\vx]$ where $\vx=(x_1, \dots, x_d)^\top$ and algebraic varieties in $\C^d$.

For a finite set of polynomials $\{f_j\}_{j=1}^m\subset \C[\vx]$, the Jacobian of  $\{f_j\}_{j=1}^m$ is the $m\times d$ matrix given by
\begin{equation}   \label{poly-jacobian}
     J(\vx) := \begin{pmatrix} \partial f_1/\partial x_1 & \cdots &\partial f_1/\partial x_d \\
                                   \vdots & \vdots & \vdots \\
                                \partial f_m/\partial x_1 & \cdots &\partial f_m/\partial x_d
               \end{pmatrix}.
\end{equation}
Let $V$ be an algebraic variety in $\C^d$ and $\vx\in V$. Assume that $\bI(V)=\left<f_1, \dots, f_m\right>$ and the Jacobian of $\{f_j\}_{j=1}^m$ is $J(\vx)$. Several results are well known. First the local dimension of $V$ around $\vx$ is $d-\min_{\vy} \rank(J(\vy))$
where $\vy$ ranges over the local analytic manifold points of $V$ arbitrarily near $\vx$. The dimension of $V$, denoted by $\dim V$, is the maximum of the local dimensions (see Definition 2.3 in \cite{Ken}). Furthermore, if $V$ is irreducible then the local dimension of $V$ is a constant, which is of course just $\dim V$.  An equivalent  definition of dimension of $V$ is defined as the Krull dimension of $\bI(V)$.

Note that a complex algebraic variety $V$ may contain real points. We use $V_\R$ to denote the real points of $V$. Assume that $\bI(V)=\left<f_1, \dots, f_m\right>$. Each $f_j$ can be written uniquely as $f_j(\vx) = g_j(\vx)+ ih_j(\vx)$ where both $g_j, h_j$ are polynomials with real coefficients. It is easy to see that $V_\R$ is the real zero locus of the real polynomials $g_1, \dots, g_m, h_1, \dots, h_m$. According to Theorem 2.3.6 in \cite{real}, any real semi-algebraic subset of $\R^d$ is homeomorphic as a semi-algebraic set to a finite disjoint union of hypercubes. Thus one can define the real dimension of $V_\R$, denoted by $\dim_\R V_\R$ as the maximal dimension of a hypercube in this decomposition. An important fact is:

\begin{lem}  \label{lem-realdim}
   Let $V$ be an algebraic  variety in $\C^d$. Then  $\dim_\R V_\R\leq \dim V$.
\end{lem}
\Proof This is already shown in Section 2.1.3 in \cite{E15} under the assumption that $V$ is defined by the locus of a collection of polynomials with real coefficients. So we only need to consider the case in which $\bI(V)=\left<f_1, \dots, f_m\right>$ and not all $f_j$ are real polynomials. Write $f_j(\vx) = g_j(\vx)+ih_j(\vx)$ where $g_j(\vx)$ and $h_j(\vx)$ are the unique polynomials with real coefficients. Then $V_\R$ is the real zero locus of the real polynomials $\{g_j(\vx), h_j(\vx)\}_{j=1}^m$. Let $W$ be the complex zero locus of $\{g_j(\vx), h_j(\vx)\}_{j=1}^m$. Then $\dim W \geq \dim_\R V_\R$. However, $W \subseteq V$ and hence $\dim W \leq \dim V$. The lemma follows.
\eproof

Almost all varieties we consider in this paper will be the zero locus of a collection of homogeneous polynomials. Any such variety can naturally be viewed as a projective variety in $\PP(\C^d)$. Let $\sigma: \C^d\setminus \{0\} \lra \,\PP(\C^d)$ be the canonical map $\sigma(\vx) = [\vx]$, where $[\vx]\in \PP(\C^d)$ denotes the line through $\vx$. We shall also often consider the {\em projectivization} of a set $S\subset \C^d\setminus \{0\}$, to be $[S]=\sigma(S)$.

\subsection{Generalized Matrix Recovery}
The aim of this subsection is to investigate the generalized matrix recovery problem introduced earlier, through the study of related algebraic varieties. Let $L_j: \H^{n} \times \H^{m} \lra \H$ be a bilinear function where $\H=\R$ or $\C$.
Suppose that $V_j\subset \H^n, j=1,\ldots,N,$ and $W\subset \H^m$ are algebraic varieties. Our objective is to show that under certain conditions an element $\vw\in W$ can be uniquely determined by a series of ``observations'' in the form of $L_j(\vx_j,\vw)$ where $\vx_j\in V_j$.
As said before, it is enough to consider whether $\{\vw\in W: L_j(\vx_j,\vw)=0, \vx_j\in V_j, j=1,\ldots,N\}$ only contains zero point.
For matrix recovery, we usually assume $V_j$ and $W$ are varieties in the space of matrices. The bilinear functions  $L_j$ is usually in the form $L_j(A, Q) = \tr(AQ)$ or more generally $L_j(A,Q)= \tr(\tau(A)Q)$ where $A, Q\in \H^{d\times d}$ and  $\tau: \H^{d\times d} \lra \H^{d\times d}$ is a linear map.

\begin{defi}   \label{defi-admissible}  {\rm
   Let $V$ be the zero locus of a collection of homogeneous polynomials in $\C^{d}$ with $\dim V>0$ and let $\ell_\alpha: \C^d\lra \C$, $\alpha\in I$, be a family of (homogeneous) linear functions where $I$ is an index set. We say $V$ is {\em admissible} with respect to $\{\ell_\alpha:\alpha \in I\}$ if  $\dim (V\cap \{\vx\in \C^d:\ell_\alpha(\vx)=0\}) <\dim V$ for all $\alpha\in I$.
}
\end{defi}

It is well known in algebraic geometry that if $V$ is irreducible in $\C^d$ then $\dim (V\cap Y) = \dim(V)-1$ for any hyperplane $Y$ that does not contain $V$ (see Corollary  4 in \cite{Coxbook}).  Thus the above admissible condition is equivalent to the property that no irreducible component of $V$ of dimension $\dim V$ is contained in any hyperplane $\ell_\alpha(\vx)=0$. In general without the irreducibility condition, admissibility is equivalent to that for a {\em generic point} $\vx\in V$ and any small neighborhood $U$ of $\vx$, $U\cap V$ is not completely contained in any hyperplane $\ell_\alpha(\vx)=0$.

We now prove the following theorem, which is one of the key theorems of this paper. It will be applied to matrix recovery and used to establish results for phase retrieval.

\begin{theo}  \label{theo-LowRank}
For $j=1, \dots, N$ let $L_j:\C^{n} \times \C^{m}\rightarrow \C$ be bilinear functions  and $V_j$ be algebraic varieties in $\C^{n}$ defined by homogeneous polynomials. Set $V := V_1\times \dots \times V_N \subseteq (\C^{n})^N$. Let $W\subset \C^{m}$ be an algebraic  variety given by homogeneous polynomials. For each fixed $j$, assume that $V_j$ is admissible with respect to the linear functions $\{f^\vw (\cdot)=L_j(\cdot,\vw):~\vw\in W\setminus \{0\}\}$.
\begin{itemize}
\item[\rm (1)]~ Assume that $N \geq \dim W$ and let $\delta:=N-\dim W+1\geq 1$. Then there exists an algebraic subvariety $Z\subset V$ with $\dim(Z) \leq \dim(V) -\delta$  such that, for any $X=(\vx_j)_{j=1}^N \in V\setminus Z$ and $\vw\in W$, $L_j(\vx_j,\vw) = 0$ for all $1 \leq j\leq N$ implies $\vw=0$.
\item[\rm (2)]~ If $N < \dim W$,  for any $X=(\vx_j)_{j=1}^N \in V$, there exists a nonzero $\vw\in W$ such that $L_j(\vx_j, \vw)=0$ for all $1 \leq  j \leq N$.
\end{itemize}
\end{theo}
\Proof We first prove (1). Define $\Phi_X:W\rightarrow \C^N$ by $\Phi_X(\vw) = (L_j(\vx_j,\vw))_{j=1}^N$. We show that for any $\vw\in W$, $\Phi_X(\vw)=0$  if and only if $\vw=0$. Let $\G$ be the subset of $ [V]\times  [W] \subset \PP((\C^{n})^N)\times \PP(\C^{m})$ such that $([X],[\vw])\in \G$ if and only if $\Phi_X(\vw) = 0$, i.e. $L_j(\vx_j,\vw)=0$ for all $j$.
Note that $\G$ is the zero locus of homogeneous polynomials $L_j(\vx_j,\vw)=0$ in the entries $X=(\vx_j)_{j=1}^N$ and $\vw$. Thus  $\G$ is a projective variety of $\PP((\C^{n})^N)\times \PP(\C^{m})$. We consider its dimension. Let $\pi_1$ and $\pi_2$ be projections from $\PP((\C^{n})^N)\times \PP(\C^{m})$ onto the first and the second coordinates, respectively, namely
$$
   \pi_1([X],[\vw])=[\vx_1,\ldots,\vx_N],\quad \pi_2([X],[\vw])=[\vw].
$$
We claim that $\pi_2(\G)=[W]$, the projectivization of $W$.
Indeed, for any fixed nonzero $\vw_0\in W$ the elements $\vx\in\C^n$ such that $L_j(\vx,\vw_0) = 0$ form a hyperplane in $\C^{n}$ with codimension 1.  It follows that this hyperplane must intersect $V_j\setminus \{0\}$ (see \cite[Prop.11.4]{alge}). Let $\vy_j \neq 0$ be in  the intersection. Set  $X_0 := (\vy_1, \dots, \vy_N)$. Then we have  $([X_0],[\vw_0]) \in \G$ and thus $\pi_2([X_0],[\vw_0])=[\vw_0]$. Consequently we have $\pi_2(\G)=[W]$.
Now  $[W]\subset \PP(\C^{m})$ is a projective variety because it is the zero locus of homogeneous polynomials. Thus
\begin{equation}\label{eq:dimeq1}
     \dim(\pi_2(\G))\,\,=\,\,\dim W-1.
\end{equation}

We next consider the dimension of the preimage $\pi_2^{-1}([\vw_0])\subset \PP((\C^{n})^N))$ for a fixed $[\vw_0]\in \PP(\C^{m})$.
Let $V'_j := V_j\cap H_j$ where $H_j:=\{\vx\in \C^{n}: L_j(\vx,\vw_0)=0\}$
is a hyperplane. The admissibility property of $V_j$ implies that  $\dim (V'_j) =\dim(V_j) -1$ (see \cite{alge}). Hence after projectivization the preimage $\pi_2^{-1}([\vw_0])$ has dimension
\begin{equation}\label{eq:dimeq2}
  \dim \pi_2^{-1}([\vw_0]) = \sum_{j=1}^N (\dim(V_j)-1)-1 = \dim(V)-N-1.
\end{equation}
By \cite[Cor.11.13]{alge}, we have
\begin{eqnarray*}
%\begin{align*}
{\rm dim}(\G) &=& {\rm dim}(\pi_2(\G))+{\rm dim}(\pi_2^{-1}([\vw_0]))\\
&=& (\dim W-1)+(\dim(V)-N-1)\\
&=& \dim(V)+\dim W-N-2
%\end{align*}
\end{eqnarray*}
where for the second equality we use (\ref{eq:dimeq1}) and (\ref{eq:dimeq2}). If $N\geq \dim W$ then
\begin{equation}\label{eq:dimpi1}
    {\rm dim}(\pi_1(\G)) \leq {\rm dim}(\G)=\dim(V)+\dim W-N-2 = \dim(V)-\delta-1.
\end{equation}
Here, we use the result that  the dimension of the projection  is less than or equal to the dimension of the original variety, see \cite[Cor.11.13]{alge}. Note that $\pi_1(\G)$ is itself a projective variety. Let $Z$ be the lift of $\pi_1(\G)$ into the vector space $(\C^{n})^N$. Then
$$
    \dim Z \,\leq \, \dim(V)-\delta.
$$
For any $X =(\vx_j)_{j=1}^n \in V\setminus Z$, by the  definition of $\G$,  $\Phi_X(\vw) = 0$ for $\vw\in W$ implies $\vw=0$.

We now prove (2), namely $\Phi_X$ cannot be injective if $N < \dim W$. Set
$$
   Z_X :=  \Bigl\{[\vw]\in \PP(\C^{m}): ~\vw \in \C^{m}, \Phi_X(\vw)=0\Bigr\}.
$$
Then $Z_X$ is a linear subspace in $\PP(\C^{m})$ with $\dim(Z_X) \geq m-1-N$. The projective variety $[W]\subseteq \PP(\C^{m})$ has dimension $\dim W-1$. If $N\leq \dim W-1$ then
$$
    \dim (Z_X)\dim ([W]) \geq  m-1,
$$
which implies that (see \cite[Prop.11.4]{alge})
$$
     Z_X \cap [W] \neq \emptyset.
$$
Thus for $N\leq \dim W-1$  there exists a non-zero $\vw_0\in \C^{m}$ with $[\vw_0]\in Z_X \cap [W] $ satisfying $\Phi_X(Q_0)=0$. It follows that $\Phi_X$ is not injective on $W$.
\eproof

\vspace{1ex}

\begin{coro}  \label{theo-LowRankReal}
Under the hypotheses of Theorem  \ref{theo-LowRank}, let $V_\R$ be the real points of $V$. Assume that $\dim_\R V_\R=\dim V$. Then there exists a real algebraic subvariety  $\tilde Z\subset V_\R$ with $\dim_\R(\tilde Z) < \dim_\R(V_\R) $ such that, for any $X=(\vx_j)_{j=1}^N \in V_\R\setminus \tilde Z$  and $\vw\in W$, $L_j(\vx_j,\vw) = 0$ for all $1 \leq j\leq N$ implies $\vw=0$.
\end{coro}
\Proof Let ${\tilde Z}:=Z_\R$ be the real points of $Z$ where
the definition of $Z$ is given in Theorem  \ref{theo-LowRank}.
 Note that
$$
   \dim_\R(\tilde Z)\leq \dim(Z)\leq \dim(V)-\delta=\dim_\R(V_\R)-\delta
$$
where $\delta = N -\dim W +1$ as in Theorem  \ref{theo-LowRank}. The Corollary now follows immediately.
\eproof

\vspace{2mm}

We now apply Theorem \ref{theo-LowRank} to study matrix recovery. In this setting we consider bilinear functions $L_j(A,Q)= \tr(AQ)$ where $A, Q\in \C^{d\times d}$.  We shall let $W= {\mathcal M}_{d,r}(\C)$ where as before
$$
  {\mathcal M}_{d,r}(\H):=\Bigl\{Q\in \H^{d\times d}: {\rm rank}(Q)\leq r\Bigr\}, \mhsp\H=\C \mbox{~or~}\R.
$$
Note that ${\rm rank}(Q)\leq r$ is equivalent to the vanishing  of all $(r+1)\times (r+1)$ minors of $Q$ and that these $(r+1)\times (r+1)$ minors are  homogeneous polynomials in the entries of $Q$.
Hence, ${\mathcal M}_{d,r}(\H)$ is an algebraic variety in $\H^{d^2}$. For $\H=\C$ it has dimension  $2dr-r^2$ \cite[Prop. 12.2]{alge}  and degree $\prod_{i=0}^{d-r-1}\frac{(d+i)!\cdot i!}{(r+i)!\cdot (d-r+i)!}$ \cite[Example 19.10]{alge}.  The projectivization of $ {\mathcal M}_{d,r}(\H)$ is a projective variety in
$\PP(\H^{d^2})$ and is called a {\em determinantal variety}. It is also well known that a determinantal variety is irreducible (see \cite{Nar86}). Theorem \ref{theo-LowRank} implies the following theorem:

\vspace{2mm}

\begin{coro}  \label{coro-LowRank}
For $j=1, \dots, N$ let $L_j: \C^{d\times d} \times \C^{d\times d}\rightarrow \C$ be bilinear functions  and $V_j$
be algebraic varieties in $\C^{d\times d}$ defined by homogeneous polynomials. Set $V := V_1\times \dots \times V_N \subseteq (\C^{d\times d})^N$.
%Let $W ={\mathcal M}_{d,r}(\C)$.
For each fixed $j$, assume that $V_j$ is admissible with respect to the linear functions
$\{f^Q(\cdot )=L_j(\cdot ,Q):~Q\in {\mathcal M}_{d,r}(\C)\setminus \{0\}\}$.
\begin{itemize}
\item[\rm (1)]~ Assume that $N \geq 2rd-r^2$,  $Q\in {\mathcal M}_{d,r}(\C)$ and set $\delta:=N-2rd-r^2+1\geq 1$. Then there exists an algebraic subvariety $Z\subset V$ with $\dim(Z) \leq \dim(V) -\delta$  such that for any $\A=(A_j)_{j=1}^N \in V\setminus Z$, $L_j(A_j,Q) = 0$ for all $1 \leq j\leq N$  implies $Q=0$.
\item[\rm (2)]~ If $N < 2rd-r^2$ then for any $\A=(A_j)_{j=1}^N \in V$ there exists a nonzero $Q\in {\mathcal M}_{d,r}(\C)$ such that $L_j(A_j, Q)=0$ for all $1 \leq j \leq N$.
\end{itemize}
\end{coro}
\Proof This follows immediately from Theorem \ref{theo-LowRank} by taking $\C^n = \C^m = \C^{d\times d}$ and $W ={\mathcal M}_{d,r}(\C)$. Here, we use  $\dim W = 2rd-r^2$ \cite[Prop. 12.2]{alge}.
\eproof

\medskip

For $W={\mathcal M}_{d,r}(\C)$  and $L_j(A_j,Q)=\tr(A_jQ)$, the hypothesis in Corollary \ref{coro-LowRank} that  $V_j$ is admissible with respect to the linear functions $\{f^Q(\cdot )=L_j(\cdot ,Q):~Q\in {\mathcal M}_{d,r}(\C)\setminus \{0\}\}$ is satisfied under many circumstances, e.g. if $V_j = {\mathcal M}_{d,r_j}(\C)$ where $r_j \geq 1$ (see the proof of Theorem \ref{th:nonsig}). In the next section more examples will be given.

\begin{coro}  \label{coro-LowRankReal}
Under the hypotheses of Corollary \ref{coro-LowRank}, suppose  that $\dim_\R V_\R=\dim V$ and $Q\in {\mathcal M}_{d,r}(\C)$.
Then there exists a real  algebraic subvariety  $\tilde Z\subset V_\R$ with $\dim_\R(\tilde Z) \leq \dim_\R(V_\R) -\delta$ where $\delta=N-(2rd-r^2)+1\geq 1$, such that for any $\A=(A_j)_{j=1}^N \in V_\R\setminus \tilde Z$, $L_j(A_j,Q) = 0$ for all $1 \leq j\leq N$  implies $Q=0$.
\end{coro}
\Proof   Let ${\tilde Z}:=Z_\R$ be the real points of $Z$, where $Z$  is the algebraic variety in  Corollary \ref{coro-LowRank}. Note that
$$
   \dim_\R(\tilde Z)\leq \dim(Z)\leq \dim(V)-\delta=\dim_\R(V_\R)-\delta.
$$
 The Corollary now follows immediately from Corollaries \ref{theo-LowRankReal} and \ref{coro-LowRank}.
\eproof

\subsection{Nonsingular Bilinear Form}
For $\H=\R$, Theorem \ref{prop-2.1} shows the equivalence between the generalized phase retrieval property and the existence of nonsingular symmetric bilinear form. Inspired by this result, we take a detour from phase retrieval to consider nonsingular bilinear form in this subsection. First we recall some notations concerning bilinear form.
Let  $\LL:\H^p\times \H^q\rightarrow \H^N$  be a bilinear form  of size $(p,q,N)$ given by $\LL(\vx,\vy)=(\vx^\top B_1\vy,\ldots,\vx^\top B_N\vy)\in \H^N$ where $\vx\in \H^p, \vy \in \H^q$ and $B_j\in \H^{p\times q}$. We call the bilinear form $\LL$ the {\em bilinear form corresponding to $B_1, \dots, B_N$}. $\LL$ is said to be {\em nonsingular} if $\LL(\vx,\vy)=0$ implies $\vx=0$ or $\vy=0$. We shall call $\LL$ a {\em real} bilinear form if $\H=\R$.

\begin{theo}\label{th:nonsig}
 Let $p,q\geq 1$, $N\geq p+q-1$ and $1\leq r_1,\ldots,r_N\leq \min\{p,q\}$. For $j=1,\ldots,N$ let $B_1,\ldots,B_N\in \H^{p\times q }$ be $N$ generic matrices with $\rank (B_j)=r_j$, where $\H=\C$ or $\R$. Then the bilinear form $\LL$ corresponding to $B_1,\ldots,B_N$ is nonsingular.
\end{theo}
\Proof We apply Theorem \ref{theo-LowRank} to prove this result.  Define
$$
  {\mathcal M}_{(q\times p),r}(\H)\,:=\,\Bigl\{Q\in \H^{q\times p}: ~\rank (Q)\leq r\Bigr\}.
$$
Then $ {\mathcal M}_{(q\times p),r}(\H)$ is an algebraic variety  and its dimension is known to be $(p+q)r-r^2$. Now
$$
  \LL(\vx,\vy)=\left(\tr(B_1\vy\vx^\top),\ldots,\tr(B_N\vy\vx^\top)\right).
$$
Hence the bilinear form $\LL$ is nonsingular if and only if
\begin{equation}\label{eq:dengnon}
 \Bigl\{Q\in \H^{q\times p}: ~\tr(B_jQ)=0, j=1,\ldots,N \Bigr\}\cap {\mathcal M}_{(q\times p),1}(\H)=\{0\}.
\end{equation}

We prove the theorem first for $\H=\C$. In Theorem \ref{theo-LowRank} we take $\C^n  = \C^{p\times q}$ and $V_j = {\mathcal M}_{(p\times q),r_j}(\C)$. Let $W ={\mathcal M}_{(q\times p), 1}(\C)$. Note that $N \geq \dim W = p+q-1$. We show that the bilinear functions $L_j(A,Q) = \tr(AQ)$ satisfies the admissibility hypothesis of Theorem \ref{theo-LowRank}. Since each $V_j$ is irreducible, we only need to show that for any nonzero $Q_0\in W$ not all $A\in V_j$ are in the hyperplane defined by $\tr(AQ_0)=0$. To see this, let $Q_0 = \vy_0\vx_0^\top$ where $\vy_0\in \C^q$ and $\vx_0\in\C^q$ are nonzero. Set $A_0 = \bar\vx_0\vy_0^* \in V_j$. Then $\tr(A_0Q_0) = \|\vx_0\|^2\|\vy_0\|^2 >0$. Thus the admissibility hypothesis is met by each $V_j$. It follows from Theorem \ref{theo-LowRank} that there exists a variety $Z\subset V := V_1\times \dots \times V_N$ with $\dim Z <\dim V$ such that  for any $(B_j)_{j=1}^N \in V\setminus Z$, (\ref{eq:dengnon}) holds, and thus $\LL$ corresponding to $B_1, \dots, B_N$ is nonsingular.  This proves the theorem for $\H=\C$.

For $\H=\R$ we notice that ${\mathcal M}_{(p\times q),r_j}(\R)$ is the real points of ${\mathcal M}_{(p\times q),r_j}(\C)$, and furthermore its real dimension is $r(p+q)-r^2$, the same as $\dim {\mathcal M}_{(p\times q),r_j}(\C)$. Thus $\dim V_\R=\dim V$. The theorem now follows directly from Corollary \ref{theo-LowRankReal}.
\eproof

\vspace{3mm}
\noindent
{\bf Remark.}
~In the complex $\H=\C$ setting part (ii) of Theorem \ref{theo-LowRank} also shows that no complex bilinear form with $N \leq p+q-2$ can be nonsingular. For the real setting $\H=\R$ the situation is quite different. We know through the above theorem that $p\#q \leq p+q-1$. But $p+q-1$ is in general not sharp. For example, as mentioned in the Introduction, for $p=q=1, 2, 4, 8$ we have $p\#q= p$. The problems of finding $p\#q$ and constructing nonsingular bilinear forms are difficult in general. Theorem \ref{th:nonsig} shows that generic $\{B_j\}_{j=1}^{p+q-1}$ with prescribed ranks $\rank (B_j)=r_j$ will yield a nonsingular bilinear form $\LL$ of size $(p,q,p+q-1)$. To our knowledge, the result is new.

We next consider the case where $N\leq p+q-2$. It is possible that a nonsingular real bilinear form of size $(p, q, N)$ still exists. A necessary condition for its existence is the Stiefel-Hopf condition (see Theorem \ref{th:hopf}). Below we provide another necessary condition. Though later we show that this condition is equivalent to the Stiefel-Hopf condition, still we decide to include it here because our proof is purely algebraic and is different from the other known proofs. We also hope the method can be helpful for finding something new.

\begin{theo}\label{th:nec}
Suppose a nonsingular real bilinear form of size $(p,q,N)$ with $N\leq p+q-2$ exists. Then the following
binomial coefficients must be even:
$$
 \binom{n}{p-1},\quad N\leq n\leq p+q-2.
$$
\end{theo}
\Proof %Without loss of generality we assume $p \leq q$.
Clearly if there is a nonsingular real bilinear of size $(p,q,N)$ exists then so does a nonsingular real bilinear of size $(p,q,n)$ for any  $n\geq N$.  Hence,  we only need to show that $ \binom{N}{p-1}$ is even.

To this end, we only need to show that if $\binom{N}{p-1}$ is odd then any real bilinear form of size $(p,q,N)$ must be singular. Assume that $\LL$ is a real bilinear form corresponding to $B_1, \dots, B_N \in \R^{p\times q}$. For any $\vy\in\C^q$ we let
$$
      Q_\vy : = (B_1\vy, B_2\vy, \dots, B_N\vy)
$$
where the columns of $Q_\vy$ are $B_1\vy, \dots, B_N\vy$. Thus $\LL$ is singular if and only if there exits a nonzero $\vy_0\in \R^q$ such that ${\rm rank}(Q_{\vy_0})\leq p-1$.

We now consider elements $(A,\vy) \in \C^{p\times N}\times \C^q$. Define the projective subvariety
$$
V_{p,N,q}:=\Bigl\{[(A,\vy)]\in \PP(\C^{p\times N}\times \C^q): ~{\rm rank}(A)\leq p-1\Bigr\}
%{\mathcal B}_{p,N}^q:=\{[Q,\vy]\in \PP(\C^{p\times N}\times \C^q): {\rm rank}(Q)\leq p-1\}.
$$
In other words, $V_{p,N,q} $ is the projectivization of the variety ${\mathcal M}_{(p\times N),p-1}(\C)\times \C^q$. Hence it has dimension
$$
   \dim V_{p,N,q} = (p-1)(p+N)-(p-1)^2 + q - 1 = N(p-1)+p+q-2.
$$
Furthermore, by \cite[Example 19.10]{alge} it has degree $\binom{N}{p-1}$.

Finally we observe that the existence of $\vy_0\in \R^q\setminus\{0\}$ such that   ${\rm rank}(Q_{\vy_0})\leq p-1$ if and only if there exists a $[(A,\vy)]\in V_{p,N,q}$ such that $\vy\in\R^q$ and the $j$-th column of $A$, say ${\mathbf a}_j$, is exactly $B_j\vy$ for each $j$. Set
$$
 {\mathcal H}:=\Bigl\{[(A,\vy)]\in \PP(\C^{p\times N}\times \C^q): ~{\mathbf a}_j-B_j\vy=0, j=1,\ldots,N\Bigr\}.
$$
Then ${\mathcal H}$ is a hyperplane in $ \PP(\C^{p\times N}\times \C^q)$ with ${\rm dim}({\mathcal H})\geq q-1$. Since $N\leq p+q-2$, we have
$$
  {\rm dim} {\mathcal H}+\dim V_{p,N,q} \geq Np+q-1= \dim(\PP(\C^{p\times N}\times \C^q))
$$
which implies $V_{p,N,q} \cap {\mathcal H} \neq \emptyset$, see \cite[Proposition 11.4]{alge}. Now all $B_j$ are real so $V_{p,N,q}$ is defined by polynomials of real coefficients. If the degree of $V_{p,N,q} = \binom{N}{p-1}$ is odd, then the intersection $V_{p,N,q} \cap {\mathcal H}$ must contain real points. Hence $\LL$ is singular. The theorem is proved.
\eproof

\vspace{2mm}
\noindent
{\bf Remark.}~~Because of symmetry, we also know that a necessary condition for the existence of nonsingular real bilinear from of size $(p,q, N)$ is that $ \binom{n}{q-1}$ is even for  all $N\leq n\leq p+q-2$. It turns out that our condition in Theorem \ref{th:nec} is equivalent to the Stiefel-Hopf condition stated in Theorem \ref{th:hopf}. To see this, we note the identity $\binom{N+t}{k}=\binom{N+t-1}{k-1}+\binom{N+t-1}{k}$. Hence by induction we have
$$
   \binom{N+t}{k}=\sum_{j=0}^t a_j\binom{N}{k-j}
$$
for some positive integers $a_j \in \N$. Assume that all the binomial coefficients  $\binom{N}{p-1},\binom{N}{p-2},\ldots,\binom{N}{N-q+1}$ are even. Then
$$
\binom{N+t}{p-1}=\sum_{j=0}^t a_j \binom{N}{p-1-j}
$$
must be even for all $t=0,\ldots, p+q-2-N$. The converse is proved by the same way.

It is worth noting that $\binom{n}{m}$ is odd if and only if the sum of $m$ and $n-m$ has no carry in base 2, i.e. the base expansion of $m$ and $n-m$ have no overlapping 1's. This fact leads to finer results on $p\#q$, which we omit here. Theses results can be found in \cite{Shapirobook}, which were obtained using different methods.

\section{Generalized Phase Retrieval With Generic Measurements}
\setcounter{equation}{0}

In this section we establish several results on the phase retrieval property of $\A=(A_j)_{j=1}^N$ where $A_j$ are chosen to be generic from some classes of matrices. The corresponding results are mostly known in the standard phase retrieval setting where all $A_j=\vf_j\vf_j^*$ with $\vf_j\in\H^d$. However, even for the standard phase retrieval the complex case is highly nontrivial. Of particular note, we show that a generic choice of $N\geq 4d-4$ subspaces (fusion frames) $\{X_j\}_{j=1}^N$ in $\C^d$ with $1 \leq \dim(X_j) \leq d-1$ have the phase retrieval property.

\begin{theo}  \label{theo-RealPRgeneric}
   Let $N\geq 2d-1$ and $1 \leq r_1, \dots, r_N \leq d$. Then a generic $\A=(A_j)_{j=1}^N\in \Herm^N(\R)$ with $\rank(A_j)=r_j$ has the phase retrieval property in $\R^d$.
\end{theo}
\Proof
By Theorem \ref{prop-2.1}, we only need show that if $Q\in {\mathcal M}_{d,1}(\R)$ and $\tr(A_jQ)=0$ for all $1 \leq j \leq N$ then $Q=0$. To prove this we apply Corollaries \ref{coro-LowRank} and \ref{coro-LowRankReal}. Set in Corollary \ref{coro-LowRank} $V = V_{r_1} \times \cdots\times V_{r_N}$, where $V_{r_j}$ denotes the symmetric determinantal variety of the set of complex symmetric matrices in $\C^{d\times d}$ with rank at most $r_j$.  The $V_{r_j}$ is  an algebraic variety which is  defined by the zero locus of a set of homogeneous polynomials. It is well known that  $\dim(V_{r_j}) = dr_j -\frac{r_j(r_j-1)}{2}$ and $\dim_\R((V_{r_j})_\R)=dr_j -\frac{r_j(r_j-1)}{2}$. Thus $\dim V = \dim_\R V_\R$.

Set $W={\mathcal M}_{d,1}(\C)$ and let $L_j(A,Q) = \tr(AQ)$. Assume we know that $V_{r_j}$ is admissible with respect to $\{f^Q(\cdot )=L_j(\cdot, Q):~Q\in {\mathcal M}_{d,1}(\C)\setminus\{0\}\}$ for all $j$ then our theorem follows immediately from Corollary \ref{coro-LowRankReal}.

Thus all it remains is to show the admissibility of $V_{r_j}$. To do so it suffices to show that at a generic point $A_0 \in V_{r_j}$ and any nonzero $Q_0\in {\mathcal M}_{d,1}(\C)$ we must have $\tr(AQ_0) \not\equiv 0$ in any small neighborhood of $A_0$ in $V_{r_j}$. If $\tr(A_0Q_0) \neq 0$ we are done. Assume that $\tr(A_0Q_0) = 0$. Write $Q_0 = \vx_0\vy_0^\top$ and apply the Tagaki factorization to $A_0$ we get
$$
      A_0 = \sum_{j=1}^s \vz_j \vz_j^\top.
$$
Now set $\hat\vz_1=\vz_1 +t\vu$ where $\vu\in\C^{d\times d}$ and let $A= \hat\vz_1\hat\vz_1^\top+\sum_{j=2}^s \vz_j \vz_j^\top$. Then
$$
     \tr(AQ) = t^2 (\vy_0^\top\vu)(\vu^\top\vx_0) + t (\vy_0^\top\vu + \vu^\top\vx_0) + \tr(A_0Q)
                = t^2 (\vy_0^\top\vu)(\vu^\top\vx_0) + t (\vy_0^\top\vu + \vu^\top\vx_0).
$$
Clearly, since $\vu$ can be arbitrary, we can pick a $\vu$ such that  $(\vy_0^\top\vu)(\vu^\top\vx_0)\neq 0$. By taking $t$ to be very small we must have $\tr(AQ_0) \not\equiv 0$ in any small neighborhood of $A_0$ in $V_{r_j}$. This completes the proof of the Theorem.
\eproof

\vspace{3mm}
\noindent
{\bf Remark.}~~Since the set of positive semidefinite matrices of rank $r$ in $\R^{d\times d}$ is an open set in $V_{r}$, where $V_{r}$ denotes the symmetric determinantal variety of the set of complex symmetric matrices in $\C^{d\times d}$ with rank at most $r$,
 Theorem \ref{theo-RealPRgeneric} also holds if we require the matrices $A_j$ be positive semi-definite.

\vspace{3mm}

It is shown in Edidin \cite{E15} that $N \geq 2d-1$ generic fusion frames have the phase retrieval property. Below we show an alternative proof using our method.

\begin{theo}[Edidin \cite{E15}] \label{theo-RealFusion}
   Let $N\geq 2d-1$ and $1 \leq r_1, \dots, r_N \leq d-1$. Then a generic set of $N$ orthogonal projection matrices $\A=(A_j)_{j=1}^N\in \Herm^N(\R)$ where $A_j^2=A_j$ and $\rank(A_j)=r_j$ has the phase retrieval property in $\R^d$.
\end{theo}
\Proof For any $s\geq 1$ let $V_s$ denote the set of complex symmetric matrices $A$ in $\C^{d\times d}$ with the property
\begin{equation}   \label{eq:ComplexProj}
     A^2 = \frac{1}{s}\tr(A) A.
\end{equation}
Clearly (\ref{eq:ComplexProj}) gives a set of homogeneous polynomial equations in the entries of $A$. Hence, $V_s$ is an algebraic variety. We next consider the dimension of $V_s$. We claim that $\rank(A)=s$ for any nonzero $A\in V_s$. To see this, $A^2=\lambda A$ where $\lambda=\frac{1}{s}\tr(A)$. Thus the eigenvalues of $A$ are $\lambda$ with multiplicity $k:=\rank(A)$ and 0 with multiplicity $d-k$. Hence $\tr(A)=\lambda k$ and $A^2 = \frac{k}{s}\lambda A$ which implies $s=k = \rank(A)$. Note also that by Jordan canonical form we easily see that $A^2=\lambda A$ can only happen if the Jordan canonical form of $A$ is diagonal. So $A$ must be diagonalizable. It follows from \cite[Theorem 4.4.13]{Matrixbook} that there exists a complex orthogonal matrix $P$ (i.e. $PP^\top=I$) such that
\begin{equation}   \label{eq:CProjdecomp}
      A=P \begin{pmatrix} \lambda I_s & 0 \\ 0 & 0 \end{pmatrix} P^\top =\lambda\sum_{j=1}^s \vv_j\vv_j^\top
\end{equation}
where $\vv_j$ is the $j$-th column of $P$ so $\{\vv_j\}_{j=1}^s$ are complex orthonormal in the sense that $\vv_i^\top\vv_j = \delta_{ij}$. Conversely, it is clear that any matrix $A$ having the form (\ref{eq:CProjdecomp}) must be in $V_s$.

Define the map $\varphi: V_s\setminus\{0\} \lra G(s,\C^d)$ by $\varphi(A)=A(\C^d)$, where
we use $A(\C^d)$ to denote the subspace $\{A\vx:\vx\in \C^d\}$ and
$G(s,\C^d)$ to  denote the Grassmannian of $s$-dimensional subspaces of $\C^d$. The map $\varphi$ is onto because any $s$-dimensional subspaces $X$ in $\C^d$ has a complex orthonormal basis $\{\vv_1, \dots,\vv_s\}$ (see \cite{Matrixbook}), and hence $X=PP^\top(\C^d)=\varphi(PP^\top)$ with $P=(\vv_1 \dots\vv_s)\in \C^{d\times s}$. Furthermore, $\varphi$ is injective on $V_s\cap \{A\in \C^{d\times d}: A^2=A\}$ because if $A_1(\C^d)=A_2(\C^d)$ then we must have $A_1=A_2R$ for some nonsingular $R\in \C^{d\times d}$. Hence $A_2A_1=A_2^2R=A_2R=A_1$ and similarly $A_2A_1 =A_2$ from $A_1R^{-1}=A_2$. Thus $A_1=A_2$.
As a consequence, $\dim(V_s\cap\{A^2=A\}) = s(d-s)$, which is the dimension of the Grassmannian. Hence $\dim V_s=s(d-s)+1$. Recall that $(V_s)_\R= V_s\cap \R^{d\times d}$. Then $(V_s)_\R\cap\{A\in \R^{d\times d}:A^2=A\}$ corresponds to the real Grassmannian $G(s,\R^d)$, which has the real dimension $s(d-s)$. Thus $\dim_\R ((V_s)_\R)=\dim(V_s)$.

Assume we know that $V_s$ is admissible with respect to $\{f^Q(\cdot)=\tr(\cdot\, Q):~Q\in {\mathcal M}_{d,1}(\C)\setminus\{0\}\}$ for all $1 \leq s \leq d-1$. We prove the theorem in exactly the same way as we have proved Theorem \ref{theo-RealPRgeneric}, namely by setting  $V = V_{r_1} \times \cdots\times V_{r_N}$, $W={\mathcal M}_{d,1}(\C)$ and let $L_j(A,Q) = \tr(AQ)$ in Corollary \ref{coro-LowRank}. Here, $V_{r_j}$ is defined by taking $s=r_j$ in $V_s$. Since each $V_s$ is just a scale multiple of a complex orthogonal projection, the theorem is equivalent to that a generic $\A=(A_j)_{j=1}^N \in V_\R$ has the phase retrieval property. The theorem thus follows immediately from Corollary \ref{coro-LowRankReal}.

Now all we need is to show the admissibility of $V_s$. The map $\varphi$ induces an isomorphism from $[V_s]$, the projectivization of $V_s$, to the Grassmannian. Since the Grassmannian is an irreducible projective variety, it follows that $V_s$ is irreducible. To show it is admissible we now only have to show that it is not contained in any hyperplane $\{A: \tr(AQ_0)= 0\}$  where $Q_0\in {\mathcal M}_{d,1}(\C)$. Write $Q_0 = \vx\vy^\top$. Then $\tr(AQ_0)=\vy^\top A\vx$. Without loss of generality we assume $y_1\neq 0$. Taking $A$ that maps $\vx$ to $\lambda {\mathbf e}_1$ for some $\lambda\neq 0$ will yield $\vy^\top A\vx = \lambda y_1 \neq 0$. This completes the proof of the Theorem.
\eproof

\begin{theo}  \label{theo-ComplexPRgeneric}
   Let $N\geq 4d-4$ and $1 \leq r_1, \dots, r_N \leq d$. Then a generic $\A=(A_j)_{j=1}^N\in \Herm^N(\C)$ with $\rank(A_j)=r_j$ has the phase retrieval property in $\C^d$.
\end{theo}
\Proof  Define a linear map $\tau: \C^{d\times d} \lra \C^{d\times d}$ by
$$
    \tau(A) = \frac{1}{2} (A+A^\top) +\frac{i}{2} (A-A^\top).
$$
It is easy to see that $\tau$ is an isomorphism on $\C^{d\times d}$ and furthermore $\tau$ restricted on $\R^{d\times d}$ is an isomorphism from $\R^{d\times d}$ to $\Herm(\C)$.
Set $L(A,Q)=\tr(\tau(A)Q)$. By Theorem \ref{prop-2.2}, it suffices to show that for a generic  $\A=(A_j)_{j=1}^N\subset \R^{d\times d}$ with $\rank(\tau(A_j))=r_j$ , if $Q\in {\mathcal M}_{d,2}(\C)$ and $L(A_j, Q)=0$ then $Q=0$.

For any $s\geq 1$ let $V_s$ denote the set of matrices $A$ in $\C^{d\times d}$ such that $\rank(\tau(A))\leq s$. The $V_s$ is clearly an algebraic variety defined by the zero locus of a set of homogeneous polynomials. Since $\tau$ is an isomorphism on $\C^{d\times d}$,  we have $\dim V_s = \dim {\mathcal M}_{d,s}(\C)=2ds-s^2$. Moreover, $\dim_\R((V_s)_\R)=\dim_\R(V_s\cap\R^{d\times d})$ is exactly the (real) dimension of the set of Hermitian matrices of rank $\leq s$, which is also $2ds-s^2$ (see also \cite[Lemma II.1]{KW15}).

We now prove the theorem in exactly the same way as before through the application of Corollaries \ref{coro-LowRank} and \ref{coro-LowRankReal}. Set $r=2$, $V = V_{r_1} \times \cdots\times V_{r_N}$, $W={\mathcal M}_{d,2}(\C)$ and let $L_j(A,Q) = L(A,Q) =\tr(\tau(A)Q)$ in Corollary \ref{coro-LowRank}.
Here, $V_{r_j}$ is defined by taking $s=r_j$ in $V_s$.
Assume that we know each $V_{r_j}$ is admissible with respect to $\{f^Q(\cdot ):=\tr(\tau(\cdot )Q)\}_{Q\in W\setminus\{0\}}$. Then the theorem follows immediately from Corollary \ref{coro-LowRankReal}, as $N \geq 4d-4 = 2rd-r^2$.

Thus all it remains is to prove the admissibility of $V_s$ for all $1 \leq s <d$. To do so it suffices to show that at a generic point $A_0 \in V_s$ and any nonzero $Q_0\in {\mathcal M}_{d,2}(\C)$ we must have $\tr(\tau(A)Q_0) \not\equiv 0$ in any small neighborhood of $A_0$ in $V_s$. Note that $\{\tau(A): A\in V_s\}={\mathcal M}_{d,s}(\C)$. Thus we only need to show that for any $B_0\in {\mathcal M}_{d,s}(\C)$ we have $\tr(BQ_0) \not\equiv 0$ in any small neighborhood of $B_0$ in ${\mathcal M}_{d,s}(\C)$.  Write
$$
   Q_0 = \vx_1\vy_1^\top+ \vx_2\vy_2^\top \shsp\mbox{where $\vx_1, \vy_1 \neq 0$}, ~ \mbox{and}\shsp   B_0 = \sum_{j=1}^s \vv_j \vu_j^\top.
$$
Let $\hat\vv_1=\vv_1 +t\vz$ and $\hat\vu_1=\vu_1 +t\vw$ where $\vz, \vw\in\C^{d\times d}$ and let $B= \hat\vv_1 \hat\vu_1^\top+\sum_{j=2}^s \vv_j \vu_j^\top$. Then
$$
     \tr(BQ_0)-\tr(B_0Q_0) = t^2 \Bigl((\vy_1^\top\vz)(\vx_1^\top\vw) +(\vy_2^\top\vz)(\vx_2^\top\vw)\Bigr) + C_0t
     =t^2 \vw^\top Q_0\vz +C_0t
$$
where $C_0\in\C$ does not depend on $t$. If $\tr(B_0Q_0)\neq 0$ we are done. Otherwise we can always find $\vz,\vw \in\C^d$ such that $\vw^\top Q_0\vz\neq 0$ because $Q_0 \neq 0$. Thus $\tr(BQ_0)-\tr(B_0Q_0) = \tr(BQ_0) \neq 0$ for sufficiently small $t$. This proves the admissibility of $V_s$.
\eproof

\vspace{3mm}
\noindent
{\bf Remark.}~~Again, since the set of positive semidefinite Hermitian matrices of rank $s$ in $\C^{d\times d}$ is an open set in the set of all Hermitian matrices of rank at most $s$, Theorem \ref{theo-ComplexPRgeneric} also holds if we require the matrices $A_j$ to be positive semi-definite.

\vspace{3mm}

We now turn to the case of complex fusion frames (projection) phase retrieval by the proving the following new result.

\begin{theo}   \label{theo-ComplexFusion}
   Let $N\geq 4d-4$ and $1 \leq r_1, \dots, r_N \leq d-1$. Then a generic set of $N$ orthogonal projection matrices $\A=(A_j)_{j=1}^N\in \Herm^N(\C)$ with $A_j^2=A_j$ and $\rank(A_j)=r_j$ has the phase retrieval property in $\C^d$.
\end{theo}
\Proof Let $\tau: \C^{d\times d} \lra \C^{d\times d}$ be
$$
    \tau(A) = \frac{1}{2} (A+A^\top) +\frac{i}{2} (A-A^\top).
$$
We have already shown it is an isomorphism on $\C^{d\times d}$ and furthermore $\tau$ restricted on $\R^{d\times d}$ is an isomorphism from $\R^{d\times d}$ to $\Herm(\C)$.

For any integer $s\geq 1$ let $V_s$ denote the set of matrices $A$ in $\C^{d\times d}$ with the property
\begin{equation}   \label{eq:ComplexProj2}
     (\tau(A))^2 = \frac{1}{s}\tr(\tau(A)) \tau(A).
\end{equation}
Note that $V_s$ is a variety in $\C^{d\times d}$. Moreover, the same arguments in the proof of Theorem \ref{theo-RealFusion} shows $\rank(\tau(A))=s$ for any nonzero $A\in V_s$, and through the Jordan Canonical Form, $\tau(A)$ is diagonalizable  which means there exists a nonsingular $P$ such that
\begin{equation}   \label{eq:ComplexProjdecomp}
      \tau(A)=P \begin{pmatrix} \lambda I_s & 0 \\ 0 & 0 \end{pmatrix} P^{-1}.
\end{equation}
Let $\tilde V_s=(V_s)_\R$ be the real points of $V_s$. For any $ A\in \tilde V_s$, $\tau(A)\in \Herm(\C)$ which implies that $\tau(A)$ is an orthogonal projection matrix.   Then $\tau$ is a one-to-one map from $\tilde V_s$ to the set of all scalar multiples of orthogonal projection matrices in $\C^{d\times d}$.  To this end,  it suffices to prove $(\tau(A_j))_{j=1}^N$ has the phase retrieval property for a generic $(A_j)_{j=1}^N\in {\tilde V}_{r_1}\times \cdots \times {\tilde V}_{r_N}$.

For the dimension of $V_s$ we compute  $\dim \tau(V_s)=\dim V_s$. Let $G(s,\C^d)$  denote  the Grassmannians of $s$-dimensional subspaces of $\C^d$. We define the map $\pi: [\tau(V_s)]\longrightarrow G(s,\C^d)\times G(d-s,\C^d)$, where $[\tau(V_s)]$ is the projectivization of $\tau(V_s)$,  by $\pi([B])=({\rm Im}(B),{\rm Ker}(B))$ for any $B\in\tau(V_s)$. If $B= \tau(A)$ has the form (\ref{eq:ComplexProjdecomp}) then it is easily checked that
$$
    {\rm Im}(B) = B(\C^d) = P(Y_s),  \mhsp {\rm Ker}(B) = P(Y_s^{\perp}),
$$
where $Y_s$ is the subspace of $\C^s\times \{0\}^{d-s}$ of $\C^d$, i.e. the $s$-dimensional subspace spanned by the first $s$ coordinates. Alternatively speaking, ${\rm Im}(B)$ is the span of the first $s$ columns of $P$ and ${\rm Ker}(B)$ is the span of the last $d-s$ columns of $P$. Since $P$ can be arbitrary, it immediately implies that the map $\pi$ is onto. We show it is also one-to-one. To see this, if there is a $Q$ such that $Q(Y_s)=P(Y_s)$ and $Q(Y_s^{\perp}) = P(Y_s^{\perp})$, it is rather straightforward to check that we must have $PR=Q$ where $R$ has the block diagonal form $R = \diag(R_1, R_2)$ with $R_1 \in \C^{s\times s}$ and $R_2 \in\C^{(d-s)\times (d-s)}$. But in this case we have
$$
    Q \begin{pmatrix} \lambda I_s & 0 \\ 0 & 0 \end{pmatrix} Q^{-1}=
    P \begin{pmatrix} \lambda I_s & 0 \\ 0 & 0 \end{pmatrix} P^{-1}.
$$
Thus $\pi$ is one-to-one. Now it follows that $\pi$ is an isomorphism and
$$
   \dim [\tau(V_s)] = \dim G(s,\C^d) +\dim G(d-s,\C^d)=2s(d-s),
$$
which yields $\dim(V_s)=\dim(\tau(V_s))=\dim([\tau(V_s)])+1=2s(d-s)+1$. This is exactly the real dimension of all real scalar multiples of  projection matrices in $\C^{d\times d}$. Thus $\dim_\R({\tilde V}_s) =\dim V_s$.

We now prove the theorem following the exactly same argument as in Theorem \ref{theo-ComplexPRgeneric}. Let  $V = V_{r_1} \times \cdots\times V_{r_N}$, $W={\mathcal M}_{d,2}(\C)$ and  $L_j(A,Q)  =\tr(\tau(A)Q)$. Here $V_{r_j}$ is defined by taking $s=r_j$ in $V_s$. Assume that we know $V_{s}$ is admissible with respect to $\{f^Q(\cdot ):=\tr(\tau(\cdot )Q)\}_{Q\in W\setminus\{0\}}$ for all $1 \leq s \leq d$. Then the theorem follows immediately from Corollary \ref{coro-LowRankReal} by taking $r=2$. Here we use
the result $\dim_\R({\tilde V}_s) =\dim V_s$.

It remains to prove the admissibility of $V_s$. To do so it suffices to show that at a generic point $A_0 \in V_s$ and any nonzero $Q_0\in {\mathcal M}_{d,2}(\C)$ we must have $\tr(\tau(A)Q_0) \not\equiv 0$ in any small neighborhood of $A_0$ in $V_s$. Note that $\{\tau(A): A\in V_s\}$ consists of all projection matrices in ${\mathcal M}_{d,s}(\C)$. Thus we only need to show that for any $B_0\in {\mathcal M}_{d,s}(\C)$ with $B_0^2=B_0$ we have $\tr(BQ_0) \not\equiv 0$ for projection matrices $B$ in any small neighborhood of $B_0$ in ${\mathcal M}_{d,s}(\C)$. Also, if $B=PCP^{-1}$ then $C^2=C$ and moreover $\tr(BQ_0)=\tr(C(P^{-1}Q_0P))$. Thus we may consider the canonical case with $B_0=J_s$ where
$$
    J_s = \begin{pmatrix}  I_s & 0 \\ 0 & 0 \end{pmatrix}\,\, \in\,\, \C^{d\times d}.
$$

Set $B_t = (I+tD)B_0(I+tD)^{-1}=(I+tD)J_s(I+tD)^{-1}$. Then all we need to show is that for some $D$ and arbitrarily small $t \neq 0$ we have $\tr(B_tQ_0) \not\equiv 0$. Since $(I+tD)^{-1} = \sum_{n=0}^\infty (-1)^n t^nD^n$, we have
$$
    \tr(B_tQ_0)   =\tr(B_0Q_0)+ \sum_{n=1}^\infty (-1)^{n-1} t^n \tr\Bigl((DJ_s-J_sD)D^{n-1}Q_0\Bigr).
$$
If there exists a $D\in \C^{d\times d}$ such that $\tr((DJ_s-J_sD)D^{n-1}Q_0)\neq 0$ for some $n \geq 1$ then we are done. For $n=1$
$$
   \tr\Bigl((DJ_s-J_sD)D^{n-1}Q_0\Bigr) = \tr\Bigl((DJ_s-J_sD)Q_0\Bigr) = \tr\Bigl(D(J_sQ_0-Q_0J_s)\Bigr).
$$

 We fist consider the case where $J_sQ_0-Q_0J_s \not\equiv 0$.  Then we can take $D=(J_sQ_0-Q_0J_s)^*$ and obtain   $\tr\Bigl(D(J_sQ_0-Q_0J_s)\Bigr)\neq 0$. We are done.
 We next only consider the case where
 $J_sQ_0-Q_0J_s\equiv 0$. If $J_sQ_0-Q_0J_s\equiv 0$ then $Q_0$ must have the form
$$
    Q_0 = \begin{pmatrix}  Q_1 & 0 \\ 0 & Q_2\end{pmatrix}
$$
where $Q_1 \in \C^{s\times s}$ and $Q_2 \in \C^{(d-s)\times (d-s)}$. Consider now $n=2$ and we have
\begin{eqnarray*}
     (DJ_s-J_sD)D^{n-1}Q_0  &=&  \begin{pmatrix}  0 & -D_{12} \\ D_{21} & 0\end{pmatrix}
                                                     \begin{pmatrix}  D_{11} & D_{12} \\ D_{21} & D_{22}\end{pmatrix}
                                                     \begin{pmatrix}  Q_1 & 0 \\ 0 & Q_2\end{pmatrix} \\
          &=& \begin{pmatrix}  -D_{12}D_{21}Q_1 &- D_{12}D_{22}Q_2 \\ D_{21}D_{11}Q_1 & D_{21}D_{12}Q_2\end{pmatrix},
\end{eqnarray*}
which yields
$$
      \tr\Bigl((DJ_s-J_sD)DQ_0\Bigr) = \tr(-D_{12}D_{21}Q_1+D_{21}D_{12}Q_2).
$$
Assume that $Q_1, Q_2 \neq 0$ then both have rank 1 because $\rank(Q_0) \leq 2$. Write $Q_1 = \vx\vy^*$ and $Q_2 = \vz\vw^*$ where $\vx,\vy\in \C^s$ and $\vz,\vw\in \C^{d-s}$. Let $\vu\neq 0$ be orthogonal to $\vz$, i.e.,  $\vz^*\vu=0$. Take $D_{12}:= \vy\vu^*$ and $D_{21} := \vu\vx^*$. Then
$$
    \tr((DJ_s-J_sD)DQ_0) =-\|\vy\|^2\|\vu\|^2\|\vx\|^2 <0.
$$
Assume one of $Q_1, Q_2$ is 0, say $Q_2=0$. Then $Q_1 \neq 0$ and $\rank(Q_1) \leq 2$. Write $Q_1 = \vx_1\vy_1^*+\vx_2\vy_2^*$ where $\vx_1, \vx_2$ are linearly independent and $\vy_1 \neq 0$. Let $\vu\in\C^{s}$ such that $\vu^*\vx_2=0$ but $\vu^*\vx_1 \neq 0$.  Set $D_{12} = \vy_1 \vz^*$ and $D_{21}=\vz\vu^*$, where $\vz\in\C^{d-s}\setminus\{0\}$. Then
$$
    \tr((DJ_s-J_sD)DQ_0) =-\|\vy_1\|^2\|\vz\|^2(\vu^*\vx_1) \neq 0.
$$
The theorem is now proved.
\eproof

\section{Minimal Measurements for Generalized Phase Retrievals}
\setcounter{equation}{0}

In this section, we focus on the question: {\em What is the minimal $N$ for which there exists an $\A =(A_j)_{j=1}^N\in \Herm^N(\H)$ having the phase retrieval property in $\H^d$,  where $\H=\R$ or $\C$?} Recall that  we use $\m_\H(d)$  to denote the minimal measurement number for which such an $\A$ with phase retrieval property in $\H^d$ exists.
%, i.e.
%$$
%  \m_\H(d):=\min\Bigl\{N:~\mbox{an $\A=(A_j)\in \Herm^N(\H)$ with the phase retrieval property in $\H^d$ exists}\Bigr\}.
%$$

It is well known that for $\H=\R$ the standard phase retrieval property always implies $N \geq 2d-1$. Thus $N=2d-1$ is sharp in this case. However, for generalized phase retrieval the situation differs considerably, and it is no longer straightforward to calculate $\m_\R(d)$. We have

\begin{theo}\label{th:rmin}
\begin{enumerate}[{\rm (i)}]
\item $\m_\R(d) \leq 2d-1$ for any odd $d$ and $\m_\R(d) \leq 2d-2$ for any even $d$.
\item For any $k\geq 1$,
$$
   \m_\R(d)= \left\{\begin{array}{cl} 2d-1,  & ~d=2^k+1\\
   2d-2, & ~d=2^k+2. \end{array}\right.
$$

\item For any $d\geq 5$,
$$
   \m_\R(d) \geq \left\{\begin{array}{cl} 2d-6\lfloor\log_2(d-1)\rfloor+6, &~~\mbox{$d$ odd}\\
                2d - 6\lfloor\log_2(d-2)\rfloor+4, &~~\mbox{$d$ even}.
                \end{array}\right.
$$
\end{enumerate}
\end{theo}
\Proof The key ingredient in the proof of this theorem is the fact that $\A=(A_j)_{j=1}^N\subset \Herm^N(\R)$ has the phase retrieval property if and only if the symmetric bilinear form corresponding to the matrices $\{A_j\}_{j=1}^N$ is nonsingular (Theorem \ref{prop-2.1}).  Furthermore, if there exists a nonsingular symmetric  bilinear form of size $(d,d,N)$ with $N>d$ then there exists an embedding (and hence an immersion) of the projective space $\PP(\R^d)=\PP^{d-1}$ in $\R^{N-1}$, see \cite[Theorem 6.3]{james}.

\noindent
(i)~~ Clearly we have $\m_\R(d) \leq 2d-1$ by Theorem \ref{theo-RealPRgeneric}. For even $d$ it is known that there exists a nonsingular symmetric bilinear form with size $(d,d,2d-2)$, see \cite{hopf40} or \cite[Page 260]{Shapirobook}. Thus $\m_\R(d) \leq 2d-2$.

\noindent
(ii)~~For the case $d=2^k+1$ we apply the result in \cite{embedding1} that for this $d$, $\PP(\R^{d})$ can not be embedded  into $\R^{2d-3}$. Thus  if $(A_j)_{j=1}^N$
has phase retrieval property, then $N-1\geq 2d-2$, which implies $\m_\R(d) \geq 2d-1$. Thus $\m_\R(d)=2d-1$ because we already know $\m_\R(d) \leq 2d-1$.

For the case $d= 2^k+2$, by (i) we have $\m_\R(d) \leq 2d-2$. It was shown in  \cite{embedding2} that for this $d$,  $\PP(\R^{d})$ can not be embedded  into $\R^{2(d-1)-2}=\R^{2d-4}$ (see also \cite[page 272]{james}). This implies $\m_\R(d)-1\geq 2d-3$.  Hence $\m_\R(d)=2d-2$.

\noindent

(iii)~~Here we use a non-immersion result of Davis \cite{davisnon} that $\PP^{2(n+\alpha(n)-1)}=\PP(\R^{2(n+\alpha(n))-1})$ can not be embedded into $\R^{4n-2\alpha(n)}$ for any $n\geq 1$, where $\alpha(n)$ denotes the number of $1$'s in the binary expansion of $n$. It follows that
\begin{equation}  \label{eq:davis}
     \m_\R(2n+2\alpha(n)-1) \geq 4n-2\alpha(n)+2.
\end{equation}
Let $\SS=\{n+\alpha(n): n\in\N\}$. Unfortunately, $\SS \neq \N$. For example, $6\not\in\SS$. Nevertheless, observe that $\alpha(n+1)=\alpha(n)+1$ for even $n$ and $\alpha(n+1) = \alpha(n)+1-k \leq \alpha(n)$ for odd $n$ where $k$ is the smallest positive integer such that $n \equiv 2^k-1 \wmod{2^k}$. Thus $n+1+\alpha(n+1)-(n+\alpha(n))\leq 2$ which implies that $\SS$ cannot miss two consecutive integers. In particular, if $m\not\in \SS$ then $m-1=n+\alpha(n)\in \SS$ for some even $n$.

We now derive a lower bound for $\m_\R(d)$. First consider odd $d=2s-1$ and $s\in\SS$ with $s=n+\alpha(n)$ for some $n$. We have $n \geq 3$ because $d\geq 5$. By (\ref{eq:davis}) we have
$$
    \m_\R(d)=\m_\R(2n+2\alpha(n)-1) \geq 4n-2\alpha(n)+2=2d-6\alpha(n)+4.
$$
Since $\alpha(n) \leq  \lfloor\log_2(n+1)\rfloor$ for all $n$ and $n =s-\alpha(n)$ we have
\begin{equation}  \label{eq:alpha}
   \log_2(n+1) \leq  \log_2\Bigl(s+1-\alpha(n)\Bigr) = \log_2\Bigl(\frac{d-1}{2}+2-\alpha(n)\Bigr).
\end{equation}
If $\alpha(n) \geq 2$ then we have $\log_2(n+1) \leq \log_2(d-1) -1$. Thus $\alpha(n) \leq  \lfloor\log_2(d-1)\rfloor-1$ and hence $\m_\R(d) \geq 2d-6\lfloor\log_2(d-1)\rfloor +10$. If $\alpha(n)=1$ then $n=2^k$ and $d=2^{k+1}+1$ with $k \geq 1$, and in this case by (iii) we actually have the stronger estimate $\m_\R(d) = 2d-1 \geq 2d-6\lfloor\log_2(d-1)\rfloor +10$.

Next consider $d=2s-1$ and $s\not\in\SS$. Thus $s-1 =n+\alpha(n)$ for some even $n$, and $d=2n+2\alpha(n)+1$. Again by (\ref{eq:davis}) we have
$$
    \m_\R(d)\geq\m_\R(2n+2\alpha(n)-1) \geq 4n-2\alpha(n)+2=2d-6\alpha(n).
$$
But $n$ is even so its last digit is 0 and hence $\alpha(n)=\alpha(n+1)-1$. Now $n+1=s-\alpha(n)$, and similar to (\ref{eq:alpha}) we have
$$
    \log_2(n+1) \leq  \log_2\Bigl(s-\alpha(n)\Bigr) = \log_2\Bigl(\frac{d-1}{2}+1-\alpha(n)\Bigr)
     \leq \log_2(d-1)-1.
$$
Hence $\m_\R(d) \geq 2d-6\lfloor\log_2(d-1)\rfloor +6$. This completes the proof of (iv) for odd $d$.

For even $d=2s$ we can apply the obvious result $\m_\R(d) \geq \m_\R(d-1)$, and the conclusion follows.
\eproof

\vspace{3mm}
\noindent
{\bf Remark.}~~  Part ({\rm ii}) in Theorem \ref{th:rmin} implies  $\m_\R(4)=6$, which answers the {\em Smoothie Problem}. In \cite{smoothie} Edidin offers a smoothie to the first person who answers the question whether there exists a fusion frame with $5$ subspaces in $\R^4$ having the phase retrieval property. Our result proves that this is impossible. In \cite{Xu15} the author has constructed a fusion frame with $6$ rank $2$ subspaces in $\R^4$ having the phase retrieval property.

\vspace{3mm}

We next present results for the complex case. These results are again obtained from known results on embedding of projective spaces. The best known lower bound for the standard  phase retrieval is $4d-3-2\alpha(d-1)+\epsilon_\alpha$ where $\alpha(d-1)$ is the number of 1's in the binary expansion of $d-1$ and $\epsilon_\alpha$ is defined below, which follows from the lower bound $4d-2-2\alpha(d-1)+\epsilon_\alpha$ for information completeness of POVMs with respect to pure states \cite{HMW13}. We prove

\begin{theo}\label{th:clower}
Let $d>4$. Then $4d-2-2\alpha+\epsilon_\alpha \leq \m_\C(d)\leq 4d-3-\alpha-\delta$,
where $\alpha = \alpha(d-1)$ denotes the number of $1$'s in the binary expansion of $d-1$,
$$
\epsilon_\alpha= \left\{\begin{array}{cl} 2  & ~d \text{ odd},\, \alpha \equiv 3 \wmod 4\\
   1 &   ~d \text{ odd}, \,\alpha \equiv 2 \wmod 4\\
   0  & ~\text{otherwise.}
   \end{array}\right.
   \shsp \mbox{and} \shsp
   \delta =  \left\{\begin{array}{cl} 0  & ~d \text{ odd}\\
   1 &   ~d \text{ even}.
   \end{array}\right.
$$
\end{theo}
\Proof
The upper bound, proved for  information completeness of POVMs with respect to pure states \cite[Theorem 3]{HMW13}, was obtained via constructions in Milgram \cite{Mil67}. Since information completeness of POVMs with respect to pure states is a special case of generalized phase retrieval in which one of the matrices $A_j$ is set to be the identity matrix, the upper bound also stands as an upper bound of $\m_\C(d)$. We remark that the upper bound actually holds for $d>2$, not just $d>4$.

We next consider the lower bound.  Assume that $\m_\C(d)\geq 3d$, and let $\A=(A_j)_{j=1}^N\subset \Herm^N(\C)$ have phase retrieval property.  Then $N\geq \m_\C(d)\geq 3d$. Define the map $\psi: \C^d\rightarrow {\mathbb S}^{N-1}(\R)$ by $\psi(\vx)=
\frac{\MM_\A(\vx)}{\|\MM_\A(\vx)\|}$, where ${\mathbb S}^{N-1}(\R)$ denotes the real $(N-1)$-dimensional unit sphere in $\R^{N}$. Note that $\A$ has the phase retrieval property. Therefore $\psi(\vx)=\psi(\vy)$ if and only if $\vx=\lambda \vy$ for some  $\lambda \in \C$ which implies that $\psi$ is a topological embedding of $\PP(\C^d)$ in ${\mathbb S}^{N-1}(\R)$.
 We recall a well-known result which says that,  the manifold $M$ can be embedded in $\R^{\dim(M)+k}$ if and only if $M$ can be embedded in ${\mathbb S}^{\dim(M)+k}$ provided  $k\geq 1$ (see \cite[Page 257]{james}).  Now observe that $N-1\geq 3d-1>\dim \PP(\C^d)$ which implies that we can construct a topological embedding of $\PP(\C^d)$ in $\R^{N-1}$.
 We  now use the following result: if there exists a topological embedding of $\PP(\C^d)$ in $\R^{N-1}$ then there exists a smooth embedding provided $N\geq 3d$  (\cite[Corollary 1.5]{james} and \cite{haefliger}).
 Hence, there exists a smooth embedding of $\PP(\C^d)$ in $\R^{N-1}$.
 But the results in \cite{mayer65} shows that $\PP(\C^d)$ can not be smoothly embedded in $\R^{4(d-1)-2\alpha+\epsilon_\alpha}$. Consequently $N-1\geq 4(d-1)-2\alpha+\epsilon_\alpha+1$ and hence $\m_\C(d)\geq 4d-2-2\alpha+\epsilon_\alpha$.

We still need to consider the case $\m_\C(d)\leq 3d-1$. If $\m_\C(d)\leq 3d-1$ we can then construct an
$\A=(A_j)_{j=1}^N\subset \Herm^N(\C)$ with $N=3d$ having the phase retrieval property because $N \geq \m_\C(d)$. But now $N \geq 3d$ so the conclusion from the above case holds, namely $N-1\geq 4d-2\alpha -3 +\epsilon_\alpha$. Now we have $4d-2\alpha -3 +\epsilon_\alpha>3d-1=N$ for $d\geq 5$. This is a contradiction.
\eproof

\vspace{2mm}

The improvement from the lower bound for the standard phase retrieval in the above theorem is useful in the case $d=2^k+1$ and $k\geq 2$. Theorem \ref{th:clower} allows us to obtain the following Corollary:

\begin{theo} \label{th:cexact}
 We have the following exact values for $\m_\C(d)$:
$$
   \m_\C(d)= \left\{\begin{array}{cl} 4d-4  & ~d=2^k+1, \,k>1\\
   4d-6 & ~d=2^k+2, \,k>1 \\  4d-5 & ~d=2^k+2^j+1, \,k>j>1\\
    4d-6 & ~d=2^k+2^j+2^l+1, \,k>j>l>1.\end{array}\right.
$$
Also, $\m_\C(2)=3$.
\end{theo}
\Proof
According to Theorem \ref{th:clower}, we examine the conditions for the equality
\begin{equation} \label{cexact}
   4d-2-2\alpha+\epsilon_\alpha = \m_\C(d) = 4d-3-\alpha-\delta
\end{equation}
to hold, where $\alpha$, $\epsilon_\alpha$ and $\delta$ are as in Theorem \ref{th:clower}. It holds if and only if $\alpha = 1+\epsilon_\alpha+\delta$. For even $d$ we have $\epsilon_\alpha=0$ and $\delta=1$ . So $\alpha = \alpha(d-1)=2$. This happens if and only if $d = 2^k+2$ where $k>1$. For odd $d$ we have $\delta=0$. Hence $\alpha = 1+ \epsilon_\alpha$. Since $\epsilon_\alpha \leq 2$ we only need to consider three cases: $(\alpha, \epsilon_\alpha) \in \{(1,0), (2,1), (3,2)\}$.

In the first case $(\alpha, \epsilon_\alpha)=(1,0)$, we have $\alpha=\alpha(d-1) = 1$ and hence $d = 2^k+1$. In the second case $(\alpha, \epsilon_\alpha)=(2,1)$, $\alpha=\alpha(d-1)=2$ and it is clear that $d = 2^k +2^j +1$ where $k>j>1$. In the third case $(\alpha, \epsilon_\alpha)=(3,2)$, $\alpha=\alpha(d-1)=3$ and $d = 2^k +2^j +2^l+1$ where $k>j>l>1$.

Finally for $\m_\C(2)$,
based on Theorem  \ref{prop-2.2}, $\A$ has the phase retrieval property if and only if the (real) Jacobian of $\MM_\A$ has (real) rank $3$ everywhere on $\C^2 \setminus\{0\}$. This immediately implies $\m_\C(2)\geq 3$. Next we show the following 3 matrices have the phase retrieval property.
Set
\[
    A_1 = \begin{pmatrix}  -1 & -1 \\ -1 & 1\end{pmatrix},\quad
     A_2 = \begin{pmatrix}  -1 & -2-i \\ -2+i & 2\end{pmatrix},\quad
     A_3 = \begin{pmatrix}  1 & 0 \\ 0 & -1\end{pmatrix}.
\]
Then
%\begin{eqnarray*}
$$
\Bigl\{Q\in \C^{2\times 2} : {\rm Tr}(A_1Q)=0, {\rm Tr}(A_2Q)=0, {\rm Tr}(A_3Q)=0\Bigr\} =
\left\{ \begin{pmatrix} 2x & x i\\ -x i& 2x \end{pmatrix}: x\in \R
\right\}
$$
%\end{eqnarray*}
The eigenvalues of the matrix
$$
  \begin{pmatrix}
2x & x i\\
-x i& 2x
\end{pmatrix}
$$
are $\lambda_1=x, \lambda_2=3x$ which have the same sign.
According to Theorem \ref{prop-2.2}, $\A=(A_1,A_2,A_3)$ has the phase retrieval property. Thus $\m_\C(2)=3$.
\eproof

We remark that the minimal measurement number for standard phase retrieval for $d=2$ is known to be $4d-4=4$, see \cite{BCMN}. The above theorem shows that generalized phase retrieval the minimal measurement number can be different.

%\nocite{*}
%\bibliography{PR-min}

\begin{thebibliography}{10}

\bibitem{phaseproj2}
Saeid Bahmanpour, Jameson Cahill, Peter~G Casazza, John Jasper, and Lindsey~M
  Woodland.
\newblock Phase retrieval and norm retrieval.
\newblock {\em arXiv preprint arXiv:1409.8266}, 2014.

\bibitem{Ra13}
Radu Balan.
\newblock Stability of phase retrievable frames.
\newblock In {\em SPIE Optical Engineering+ Applications}, pages
  88580H--88580H. International Society for Optics and Photonics, 2013.

\bibitem{BCE06}
Radu Balan, Pete Casazza, and Dan Edidin.
\newblock On signal reconstruction without phase.
\newblock {\em Applied and Computational Harmonic Analysis}, 20(3):345--356,
  2006.

\bibitem{BCMN}
Afonso~S Bandeira, Jameson Cahill, Dustin~G Mixon, and Aaron~A Nelson.
\newblock Saving phase: Injectivity and stability for phase retrieval.
\newblock {\em Applied and Computational Harmonic Analysis}, 37(1):106--125,
  2014.

\bibitem{real}
Jacek Bochnak, Michel Coste, and Marie-Fran{\c{c}}oise Roy.
\newblock {\em Real algebraic geometry}, volume~36.
\newblock Springer Science \& Business Media, 2013.

\bibitem{bodmann}
Bernhard~G Bodmann and Nathaniel Hammen.
\newblock Stable phase retrieval with low-redundancy frames.
\newblock {\em Advances in computational mathematics}, 41(2):317--331, 2015.

\bibitem{phaseproj1}
Jameson Cahill, Peter~G Casazza, Jesse Peterson, and Lindsey Woodland.
\newblock Phase retrieval by projections.
\newblock {\em arXiv preprint arXiv:1305.6226}, 2013.

\bibitem{ROP}
T~Tony Cai, Anru Zhang.
\newblock Rop: Matrix recovery via rank-one projections.
\newblock {\em The Annals of Statistics}, 43(1):102--138, 2015.

\bibitem{CESV12}
E.J. Candes, Y.~Eldar, T.~Strohmer, and V.~Voroninski.
\newblock Phase retrieval via matrix completion.
\newblock {\em SIAM Journal on Imaging Sciences}, 6(1):199--225, 2013.

\bibitem{CSV12}
E.J. Candes, T. Strohmer, and V. Voroninski.
\newblock Phaselift: Exact and stable signal recovery from magnitude
  measurements via convex programming.
\newblock {\em Communications on Pure and Applied Mathematics},
  66(8):1241--1274, 2013.

\bibitem{CEHV15}
Aldo Conca, Dan Edidin, Milena Hering, and Cynthia Vinzant.
\newblock An algebraic characterization of injectivity in phase retrieval.
\newblock {\em Applied and Computational Harmonic Analysis}, 38(2):346--356,
  2015.

\bibitem{Coxbook}
David Cox, John Little, and Donal O'shea.
\newblock {\em Ideals, varieties, and algorithms}, volume~3.
\newblock Springer, 1992.

\bibitem{davisnon}
Donald~M Davis.
\newblock A strong non-immersion theorem for real projective spaces.
\newblock {\em Annals of Mathematics}, 120(3):517--528, 1984.

\bibitem{hopf}
Daniel Dugger and Daniel~C Isaksen.
\newblock The hopf condition for bilinear forms over arbitrary fields.
\newblock {\em Annals of mathematics}, pages 943--964, 2007.

\bibitem{smoothie}
Dan Edidin.
\newblock Fusion frame phase retrieval.
\newblock {\em Workshop on Frames and Algebraic \& Combinatorial Geometry,
  Universit$\ddot{a}$t Bremen, Bremen, Germany}, 2015.

\bibitem{E15}
Dan Edidin.
\newblock Projections and phase retrieval.
\newblock {\em Applied and Computational Harmonic Analysis}, 2015.

\bibitem{uniq}
Yonina~C Eldar, Deanna Needell, and Yaniv Plan.
\newblock Uniqueness conditions for low-rank matrix recovery.
\newblock {\em Applied and Computational Harmonic Analysis}, 33(2):309--314,
  2012.

\bibitem{FMW14}
Matthew Fickus, Dustin~G Mixon, Aaron~A Nelson, and Yang Wang.
\newblock Phase retrieval from very few measurements.
\newblock {\em Linear Algebra and its Applications}, 449:475--499, 2014.

\bibitem{haefliger}
Andr{\'e} Haefliger and Arnold Shapiro.
\newblock Plongements diff{\'e}rentiables dans le domaine stable.
\newblock {\em Commentarii Mathematici Helvetici}, 37(1):155--176, 1962.

\bibitem{alge}
Joe Harris.
\newblock {\em Algebraic geometry: a first course}, volume 133.
\newblock Springer Science \& Business Media, 2013.

\bibitem{HMW13}
Teiko Heinosaari, Luca Mazzarella, and Michael~M Wolf.
\newblock Quantum tomography under prior information.
\newblock {\em Communications in Mathematical Physics}, 318(2):355--374, 2013.

\bibitem{hopf40}
Heinz Hopf.
\newblock Systeme symmetrischer bilinearformen und euklidische modelle der
  projektiven r{\"a}ume.
\newblock In {\em Selecta Heinz Hopf}, pages 107--118. Springer, 1964.

\bibitem{Matrixbook}
Roger~A Horn and Charles~R Johnson.
\newblock {\em Matrix analysis}.
\newblock Cambridge university press, 2012.

\bibitem{james}
IM~James.
\newblock Euclidean models of projective spaces.
\newblock {\em Bulletin of the London Mathematical Society}, 3(3):257--276,
  1971.

\bibitem{KW15}
Michael Kech and Michael Wolf.
\newblock Quantum tomography of semi-algebraic sets with constrained
  measurements.
\newblock {\em arXiv preprint arXiv:1507.00903}, 2015.

\bibitem{Ken}
Keith Kendig.
\newblock {\em Elementary algebraic geometry}, volume~44.
\newblock Springer Science \& Business Media, 2012.

\bibitem{Lam}
Kee~Yuen Lam.
\newblock Some new results on composition of quadratic forms.
\newblock {\em Inventiones mathematicae}, 79(3):467--474, 1985.

\bibitem{embedding2}
Mark Mahowald.
\newblock On the embeddability of the real projective spaces.
\newblock {\em Proceedings of the American Mathematical Society},
  13(5):763--764, 1962.

\bibitem{mayer65}
Karl~Heinz Mayer.
\newblock Elliptische differentialoperatoren und ganzzahligkeitss{\"a}tze
  f{\"u}r charakteristische zahlen.
\newblock {\em Topology}, 4(3):295--313, 1965.

\bibitem{Mil67}
R~James Milgram.
\newblock Immersing projective spaces.
\newblock {\em Annals of Mathematics}, pages 473--482, 1967.

\bibitem{Nar86}
Himanee Narasimhan.
\newblock The irreducibility of ladder determinantal varieties.
\newblock {\em Journal of Algebra}, 102(1):162--185, 1986.

\bibitem{embedding1}
Franklin~P Peterson.
\newblock Some non-embedding problems.
\newblock {\em Bol. Soc. Mat. Mexicana (2)}, 2:9--15, 1957.

\bibitem{Shapirobook}
Daniel~B Shapiro.
\newblock {\em Compositions of quadratic forms}, volume~33.
\newblock Walter de Gruyter, 2000.

\bibitem{surv}
DB~Shapiro.
\newblock Products of sums of squares.
\newblock {\em Expo. Math}, 2:235--261, 1984.

\bibitem{steer70}
B~Steer.
\newblock On the embedding of projective spaces in euclidean space.
\newblock {\em Proceedings of the London Mathematical Society}, 3(3):489--501,
  1970.

\bibitem{V15}
Cynthia Vinzant.
\newblock A small frame and a certificate of its injectivity.
\newblock {\em arXiv preprint arXiv:1502.04656}, 2015.

\bibitem{WaXu14}
Yang Wang and Zhiqiang Xu.
\newblock Phase retrieval for sparse signals.
\newblock {\em Applied and Computational Harmonic Analysis}, 37(3):531--544,
  2014.

\bibitem{ROP1}
Chris~D White, Sujay Sanghavi, and Rachel Ward.
\newblock The local convexity of solving systems of quadratic equations.
\newblock {\em arXiv preprint arXiv:1506.07868}, 2015.

\bibitem{Xu15}
Zhiqiang Xu.
\newblock The minimal measurement number for low-rank matrices recovery.
\newblock {\em arXiv preprint arXiv:1505.07204}, 2015.

\end{thebibliography}
%\bibliographystyle{plain}

\end{document}